\documentclass[preprint,12pt]{elsarticle}
\usepackage{amsmath,amssymb,amsthm,bm,mathrsfs}
\usepackage{mathtools,graphicx,booktabs,enumitem,xcolor,microtype,float,placeins}
\graphicspath{{figures/}}
\biboptions{sort&compress}
\usepackage[hidelinks]{hyperref}
\usepackage[nameinlink,noabbrev]{cleveref}
\usepackage[margin=1in]{geometry}
\usepackage{needspace}
\usepackage{tikz}
\usetikzlibrary{arrows.meta,calc}

\newcommand{\bB}{\boldsymbol B}
\newcommand{\bb}{\widehat{\boldsymbol b}}
\newcommand{\bx}{\boldsymbol x}
\newcommand{\bv}{\boldsymbol v}
\newcommand{\bE}{\boldsymbol E}
\newcommand{\bn}{\boldsymbol n}
\newcommand{\nuin}{\nu_{in}}
\newcommand{\Om}{\Omega_i}
\newcommand{\vperp}{v_\perp}
\newcommand{\vpar}{v_\parallel}
\newcommand{\gyro}{\vartheta}
\newcommand{\dt}{\Delta t}
\newcommand{\PH}{\mathrm{PH}}
\newtheorem{theorem}{Theorem}[section]
\numberwithin{equation}{section}

\newcommand{\grad}{\nabla}
\newcommand{\Ord}{\mathcal{O}}
\crefformat{appendix}{#2#1#3}
\Crefformat{appendix}{#2#1#3}
\allowdisplaybreaks[1]

\patchcmd{\pprintMaketitle}
  {\footnotesize\itshape\elsaddress\par\vskip36pt}
  {\footnotesize\itshape\elsaddress\par\vskip24pt}
  {}{\PackageError{PH-UGKS}{Could not adjust the title-page spacing}{}}
\patchcmd{\pprintMaketitle}
  {\unvbox\keybox\par\vskip10pt}
  {\unvbox\keybox\par\nobreak\vskip10pt}
  {}{\PackageError{PH-UGKS}{Could not attach the separator to the keywords}{}}

\journal{Journal of Computational Physics}

\begin{document}
\begin{frontmatter}

\title{A polar-harmonic unified gas-kinetic scheme for magnetized ion dynamics
from cyclotron kinetics to the Hall--Pedersen constitutive limit}

\author[hkustmath]{Yixiao Wang}
\ead{ywangyp@connect.ust.hk}
\author[hkustmath]{Zhigang Pu}
\ead{zpuac@connect.ust.hk}
\author[xjtu]{Xing Ji}
\ead{jixing@xjtu.edu.cn}
\author[hkustmath,hkustsz]{Kun Xu\corref{cor1}}
\ead{makxu@ust.hk}
\cortext[cor1]{Corresponding author.}

\address[hkustmath]{Department of Mathematics, Hong Kong University of Science
and Technology, Clear Water Bay, Kowloon, Hong Kong}
\address[hkustsz]{Shenzhen Research Institute, Hong Kong University of Science
and Technology, Shenzhen, China}
\address[xjtu]{Shaanxi Key Laboratory of Environment and Control for Flight
Vehicle, Xi'an Jiaotong University, Xi'an, China}

\begin{abstract}
Magnetized ion transport in weakly ionized plasmas ranges from gyroangle-dependent kinetics to Hall--Pedersen drift--diffusion as collisionality and magnetization vary. We develop a polar-harmonic unified gas-kinetic scheme (PH-UGKS) for the ion Vlasov--BGK equation in a uniform magnetic field. The scheme evolves the full ion distribution by coupling a conservative density update to exponential evolution of its nonequilibrium component. Exact collision--rotation integration in gyroangle Fourier space is combined with a time-averaged kinetic flux that incorporates spatial transport and electric acceleration, together with a compact Hall--Pedersen correction to the density flux. The scheme conserves ion number, and analysis establishes second-order temporal consistency and asymptotic preservation of the Hall--Pedersen density limit at fixed magnetization. Numerical tests reproduce ion Bernstein dispersion and Dory--Guest--Harris growth rates and resolve changes in the gyroharmonic spectrum as the collision-to-gyrofrequency ratio varies. The driven ion-flux response agrees with an independent characteristic--Volterra reference, including finite-frequency departures from the instantaneous Hall--Pedersen relation. In collisional tests, accurate responses are obtained with time steps far larger than both the collision time and the gyroperiod. Fixed-resolution density tests confirm convergence to the corresponding Hall--Pedersen discretization. The same kinetic formulation thus connects kinetic response and macroscopic transport without switching to a fluid solver or subcycling microscopic time scales.
\end{abstract}

\begin{keyword}
unified gas-kinetic scheme \sep ion--neutral collisions \sep
ion Bernstein waves \sep Hall--Pedersen mobility
\end{keyword}

\end{frontmatter}
\thispagestyle{empty}

\section{Introduction}
\label{sec:intro}

Cross-field ion transport in weakly ionized plasmas is governed by gyromotion, ion--neutral collisions, electric acceleration, and pressure gradients~\cite{Banks1966,SchunkNagy2009}.
Non-Maxwellian ion velocity distributions observed in the high-latitude ionosphere motivate a kinetic description~\cite{StMauriceSchunk1979}.
In the near-equilibrium, long-wavelength, and low-frequency limit, the perpendicular ion flux approaches the Hall--Pedersen drift--diffusion relation~\cite{SchunkNagy2009,Pandey2024}.
Between the gyroangle-resolved kinetic regime and this limit, collisions modify the gyroharmonic spectrum of the distribution while finite-Larmor-radius effects persist.

A scheme that spans both regimes faces two requirements when the collision time and the gyroperiod are short compared with the transport time.
The first is the time-step restriction: explicit treatment of either fast process, relaxation or gyromotion, limits the admissible step to the shorter of these scales.
Rotating-grid semi-Lagrangian methods remove the gyromotion restriction by advancing the distribution in a velocity frame that rotates at the cyclotron frequency~\cite{Schild2025}.
Exponential integrators remove the restriction imposed by a stiff linear operator through exact treatment of its propagator~\cite{CoxMatthews2002,HochbruckOstermann2010}, and have been applied to highly oscillatory Vlasov equations with time steps larger than the oscillation period~\cite{FrenodHirstoagaSonnendrucker2015}.
Exponential Runge--Kutta methods treat collisional relaxation in the same way, integrating the BGK relaxation operator exactly~\cite{DimarcoPareschi2011}.

The second requirement is asymptotic consistency: at fixed numerical resolution, the scheme must reduce to a consistent discretization of the macroscopic equation as the fast scales vanish, which is the asymptotic-preserving (AP) property~\cite{Jin1999,DegondDeluzet2017}.
When a rapidly oscillating electric field or gyromotion is the only fast process, a two-scale reformulation yields AP schemes for a charged-particle beam in such a field~\cite{Crouseilles2013}, and semi-implicit particle-in-cell methods for a strong external magnetic field recover the guiding-center equation~\cite{FilbetRodrigues2016}.
When collisional relaxation is the only fast process, exponential Runge--Kutta methods recover the Euler limit of the Boltzmann equation~\cite{LiPareschi2014}, and micro--macro decompositions recover the diffusion limit of linear kinetic equations~\cite{LemouMieussens2008} and the drift--diffusion limit of collisional Vlasov--Poisson equations~\cite{CrouseillesLemou2011}.
When both processes are fast, the limit depends on their relative scaling.
In the collisional finite-Larmor-radius regime the oscillatory and dissipative limits do not commute, and multiscale particle-in-cell integrators have been constructed and analyzed in each regime~\cite{Crestetto2023}.
In the Hall--Pedersen limit both processes become fast at a fixed ion magnetization $\beta=\Om/\nuin$, with $\Om$ and $\nuin$ the ion gyrofrequency and ion--neutral collision frequency.
The Pedersen and Hall fluxes then both depend on $\beta$, so neither process is a perturbation of the other.
The present work targets this limit at fixed $\beta$ with an Eulerian gyroharmonic discretization.

The unified gas-kinetic scheme (UGKS) provides the starting point, since it addresses both requirements for collisional transport through the numerical flux.
The flux is built from a local time-dependent kinetic solution that couples transport and relaxation over the step, so that the time step is not restricted by the collision time~\cite{XuHuang2010,HuangXuYu}, and for linear kinetic models the same flux yields a consistent discretization of the diffusion limit~\cite{Mieussens2013}.
This asymptotic property persists for nonisotropic scattering and Fokker--Planck collision operators~\cite{CrouseillesMathiaudMieussens2026}, and the framework has been extended to multicomponent and partially ionized plasmas~\cite{LiuXuPlasma2017,PuXu2025} and, in its discrete formulation, to electrostatic and electromagnetic plasma models~\cite{LiuPRE2020,LiuCPC2020,LiuCaoLapentaKeppens2026}.
For magnetized ions the same construction must also carry the gyromotion over the step.
The Hall--Pedersen limit places a further demand on the flux.
The physical flux decreases as $\nuin^{-1}$, but the Maxwellian jump dissipation of an upwind flux does not.
On a fixed spatial grid this dissipation can therefore exceed the physical flux, and exact integration of relaxation and rotation does not remove it.
The numerical problem addressed here is to recover a consistent discretization of the Hall--Pedersen density equation from a kinetic flux that also retains the finite-frequency response away from the limit.

To this end, we develop a polar-harmonic unified gas-kinetic scheme (PH-UGKS) for the ion Vlasov--BGK equation~\cite{BGK} in a uniform magnetic field.
The method couples a conservative density update to exponential evolution of the nonequilibrium component of the distribution.
The Maxwellian projection and the gyroangle Fourier representation yield an exact collision--rotation propagator.
Spatial transport and electric acceleration enter through a cellwise Duhamel approximation to the time-averaged distribution used for interface flux reconstruction.
The flux retains upwinding of the nonequilibrium component and attenuates the Maxwellian jump dissipation, and in the macroscopic limit a density-flux correction replaces the leading cell-centered Hall--Pedersen contribution by a compact face flux.

The resulting scheme removes the time-step stability restrictions imposed by explicit treatment of collisional relaxation and gyromotion, conserves discrete ion number on periodic domains, and is second-order consistent in time for fixed collision and gyrofrequencies.
In the collisionless limit, its spatial flux reduces to full-distribution upwinding, whereas in the Hall--Pedersen scaling the density update converges, at fixed spatial and velocity grids and fixed slow-time step, to the compact Hall--Pedersen--Heun scheme for well-prepared perpendicular dynamics under the stated regularity and discrete moment conditions.
Numerical tests reproduce ion Bernstein dispersion and Dory--Guest--Harris growth rates in the collisionless regime and confirm convergence to the same-grid Hall--Pedersen--Heun solution.
At finite collisionality, the computed ion-flux response agrees with an independent characteristic--Volterra reference, including finite-frequency departures from the instantaneous Hall--Pedersen relation.
This agreement extends to strongly collisional driven cases with time steps exceeding both the collision time and the gyroperiod, without subcycling.
Finite-amplitude self-consistent calculations examine how collisions modify the density modes generated outside the initial Fourier support.

Section~2 introduces the kinetic model and its Hall--Pedersen limit.
Section~3 presents the scheme, and Section~4 analyzes its temporal consistency and asymptotic limits.
Numerical results and conclusions follow in Sections~5 and~6.
Appendix~A contains the Hall--Pedersen limit proof, and Appendix~B provides additional collisionless benchmarks.
\section{Magnetized ion model and Hall--Pedersen limit}
\label{sec:model}

\subsection{Gyroharmonic formulation of the Vlasov--BGK equation}
\label{sec:model-eq}

We consider a single positively charged ion species in a weakly ionized,
magnetized plasma. Collisions with the neutral background dominate the ion
relaxation on the scales of interest, and ion--ion collisions are neglected.
Ion--neutral scattering and resonant charge exchange are approximated by
constant-frequency relaxation toward a stationary neutral Maxwellian
\cite{Banks1966,SchunkNagy2009}. In the uniform magnetic field
$\bB_0=B_0\bb$, where $B_0>0$ and $\bb$ is a constant unit vector, the ion distribution
$f=f(\bx,\bv,t)$ satisfies the Vlasov--BGK equation~\cite{BGK}
\begin{equation}
\label{eq:model}
\partial_t f
+\bv\cdot\nabla_{\bx}f
+\frac{q_i}{m_i}
 \left(\bE+\bv\times\bB_0\right)\cdot\nabla_{\bv}f
=
\nuin\left(nM_n-f\right).
\end{equation}
Here $q_i$ and $m_i$ are the ion charge and mass,
$n(\bx,t)=\int_{\mathbb R^3}f(\bx,\bv,t)\,\mathrm d\bv$ is the ion
number density, and $\nuin$ is the ion--neutral collision frequency. The electric field is either
prescribed or determined self-consistently from an electrostatic closure. The
neutral Maxwellian is
\[
M_n(\bv)
=
(2\pi\theta_n)^{-3/2}
\exp\!\left(-\frac{|\bv|^2}{2\theta_n}\right),
\qquad
\theta_n=\frac{k_{\mathrm B}T_n}{m_i},
\]
where $T_n$ is the neutral temperature. Let
$v_{\mathrm{th}}=\sqrt{\theta_n}$ denote the thermal-speed scale of $M_n$,
and define the corresponding ion gyroradius by $\rho_i=v_{\mathrm{th}}/\Om$,
where $\Om=q_iB_0/m_i$ is the ion gyrofrequency.

Taking $\bb=\boldsymbol e_z$, write
$\bv=(\vperp\cos\gyro,\vperp\sin\gyro,\vpar)$, with
$\mathrm d\bv=\vperp\,\mathrm d\vperp\,\mathrm d\gyro\,\mathrm d\vpar$.
In these coordinates, the magnetic term becomes
$-\Om\partial_\gyro f$. We expand the
distribution in gyroangle Fourier modes,
\begin{equation}
\label{eq:fourier}
\begin{gathered}
f(\bx,\vperp,\gyro,\vpar,t)
=
\sum_{m\in\mathbb Z}
f_m(\bx,\vperp,\vpar,t)\,\mathrm e^{\mathrm i m\gyro},
\\
f_m
=
\frac{1}{2\pi}
\int_0^{2\pi}
f\,\mathrm e^{-\mathrm i m\gyro}\,\mathrm d\gyro,
\qquad
f_{-m}=f_m^*.
\end{gathered}
\end{equation}
Let $\boldsymbol a=(q_i/m_i)\bE$,
$a_\parallel=\boldsymbol a\cdot\bb$,
$\partial_\pm=\partial_x\pm\mathrm i\partial_y$, and
$a_\pm=a_x\pm\mathrm i a_y$. The perpendicular transport and acceleration
operators can be written as
\[
\begin{aligned}
v_x\partial_x+v_y\partial_y
&=\frac{\vperp}{2}
\left(\mathrm e^{\mathrm i\gyro}\partial_-
     +\mathrm e^{-\mathrm i\gyro}\partial_+\right),
\\
a_x\partial_{v_x}+a_y\partial_{v_y}
&=\frac{a_-\mathrm e^{\mathrm i\gyro}}{2}
\left(\partial_{\vperp}+\frac{\mathrm i}{\vperp}\partial_\gyro\right)
 +\frac{a_+\mathrm e^{-\mathrm i\gyro}}{2}
\left(\partial_{\vperp}-\frac{\mathrm i}{\vperp}\partial_\gyro\right).
\end{aligned}
\]
The factors $\mathrm e^{\pm\mathrm i\gyro}$ shift the harmonic
index by $\pm1$. Using
$\partial_\gyro\mathrm e^{\mathrm i m\gyro}
=\mathrm i m\mathrm e^{\mathrm i m\gyro}$,
the $m$th Fourier coefficient of \cref{eq:model} satisfies
\begin{equation}
\label{eq:modal-hierarchy}
\begin{aligned}
&\partial_t f_m-\mathrm i m\Om f_m
 +\vpar\partial_z f_m+a_\parallel\partial_{\vpar}f_m
 +\frac{\vperp}{2}
  \left(\partial_-f_{m-1}+\partial_+f_{m+1}\right)
\\
&\quad
 +\frac{a_-}{2}
  \left(\partial_{\vperp}-\frac{m-1}{\vperp}\right)f_{m-1}
 +\frac{a_+}{2}
  \left(\partial_{\vperp}+\frac{m+1}{\vperp}\right)f_{m+1}
\\
&\quad=\nuin\left(nM_n\delta_{m0}-f_m\right).
\end{aligned}
\end{equation}
Magnetic rotation, parallel dynamics, and ion--neutral relaxation
preserve the harmonic index, with the Maxwellian gain term
contributing only to $m=0$.

With the ion number flux
$\boldsymbol\Gamma=\int_{\mathbb R^3}\bv f\,\mathrm d\bv$,
the density and perpendicular flux components are determined by
$f_0$ and $f_{\pm1}$, respectively:
\[
\begin{aligned}
n
&=
2\pi\int_{-\infty}^{\infty}\int_0^\infty
f_0\,\vperp\,\mathrm d\vperp\,\mathrm d\vpar,
\\
\Gamma_x\pm\mathrm i\Gamma_y
&=
2\pi\int_{-\infty}^{\infty}\int_0^\infty
\vperp^2 f_{\mp1}\,\mathrm d\vperp\,\mathrm d\vpar.
\end{aligned}
\]

\subsection{Hall--Pedersen limit}
\label{sec:model-regimes}

Define $\Pi f=M_n\int_{\mathbb R^3}f\,\mathrm d\bv$, $Q=I-\Pi$, and
$h=Qf$. Then $f=nM_n+h$ and
$\boldsymbol\Gamma=\int_{\mathbb R^3}\bv h\,\mathrm d\bv$. With
$\mathscr A=\bv\cdot\nabla_{\bx}+\boldsymbol a\cdot\nabla_{\bv}$,
\cref{eq:model} becomes
\begin{equation}
\label{eq:micro-macro-system}
\begin{gathered}
\partial_t n+\nabla_{\bx}\cdot\boldsymbol\Gamma=0,
\\
\partial_t h
+Q\mathscr A(nM_n+h)
-\Om\partial_\gyro h
=-\nuin h.
\end{gathered}
\end{equation}

For the asymptotic analysis, length, velocity, time, acceleration, and
distribution are scaled by $L_0$, $v_0$, $L_0/v_0$, $v_0^2/L_0$, and
$n_0/v_0^3$, respectively. Frequencies are scaled by $v_0/L_0$, and the
same notation is retained for the dimensionless variables. For perpendicular
dynamics, set $\partial_z f=0$ and $a_\parallel=0$. We introduce
\begin{equation}
\label{eq:hp-scaling}
\varepsilon=\nuin^{-1},
\qquad
\beta=\frac{\Om}{\nuin},
\qquad
\tau=\varepsilon t.
\end{equation}
Holding the ion magnetization parameter $\beta\ge0$ fixed as
$\varepsilon\to0$ gives
$\nuin=\varepsilon^{-1}$ and $\Om=\beta/\varepsilon$, under which
\cref{eq:model} becomes
\[
\varepsilon^2\partial_\tau f
+\varepsilon\left(
 \bv_\perp\cdot\nabla_\perp f
 +\boldsymbol a_\perp\cdot\nabla_{\bv_\perp}f
 \right)
+\left(Q-\beta\partial_\gyro\right)f
=0.
\]
Let $g$ denote the leading nonequilibrium coefficient in
\[
f=nM_n+\varepsilon g+\mathcal O(\varepsilon^2),
\qquad
\int_{\mathbb R^3}g\,\mathrm d\bv=0.
\]
The terms of order $\varepsilon$ give
\begin{equation}
\label{eq:leading-nonequilibrium}
\left(I-\beta\partial_\gyro\right)g
=
\frac{M_n}{\theta_n}\,
\bv_\perp\cdot\boldsymbol d_\perp,
\qquad
\boldsymbol d_\perp
=n\boldsymbol a_\perp-\theta_n\nabla_\perp n.
\end{equation}
Let
$\boldsymbol j_\perp=\int_{\mathbb R^3}\bv_\perp g\,\mathrm d\bv$.
Taking the perpendicular first moment of \cref{eq:leading-nonequilibrium} gives
\begin{equation}
\label{eq:hp-closure}
\boldsymbol j_\perp+\beta\bb\times\boldsymbol j_\perp
=\boldsymbol d_\perp,
\qquad
\boldsymbol j_\perp
=
\frac{\boldsymbol d_\perp-\beta\bb\times\boldsymbol d_\perp}
     {1+\beta^2}.
\end{equation}
Since $\boldsymbol\Gamma_\perp=\varepsilon\boldsymbol j_\perp
+\mathcal O(\varepsilon^2)$, the leading density equation on the slow time
scale is
\begin{equation}
\label{eq:hp-density-limit}
\partial_\tau n
+\nabla_\perp\cdot
\left[
\frac{
 n\boldsymbol a_\perp-\theta_n\nabla_\perp n
 -\beta\bb\times
 \left(n\boldsymbol a_\perp-\theta_n\nabla_\perp n\right)
}{1+\beta^2}
\right]
=0.
\end{equation}
Equation~\eqref{eq:hp-density-limit} is the Hall--Pedersen drift--diffusion
limit. The Pedersen flux is aligned with $\boldsymbol d_\perp$, while the
Hall flux is directed along $-\bb\times\boldsymbol d_\perp$
\cite{SchunkNagy2009,Pandey2024}.

\section{Semigroup-integrated PH-UGKS}
\label{sec:scheme}

The semigroup-integrated polar-harmonic unified gas-kinetic scheme
(PH-UGKS) follows the UGKS principle of coupling transport and relaxation
in the numerical flux over a time step~\cite{XuHuang2010}.
The collision--rotation propagator is available in closed form in each
cell (\cref{sec:method-stage}). A Duhamel approximation built on this
propagator provides the time-averaged distribution for face
reconstruction, in place of the local characteristic integral solution
used at interfaces in classical UGKS, and the density flux carries a
compact Hall--Pedersen correction (\cref{sec:method-interface}). The construction retains full-distribution
spatial upwinding at $\nuin=0$ and recovers the compact
Hall--Pedersen--Heun density update in the scaling of
\cref{sec:hp-property}.

\subsection{Projected collision--rotation integration}
\label{sec:method-stage}

Let $K$ denote a spatial cell. The retained gyroharmonics are indexed by
$m=-M,\ldots,M$, with $M\ge1$, and the $(\vperp,\vpar)$ cells by
$(\ell,r)$. For a discrete gyroharmonic state
$q_K=\{q_{K,m,\ell,r}\}$, define
\begin{equation}
\label{eq:discrete-projection}
\langle q_K\rangle_h
=2\pi\sum_{\ell,r}\omega_{\ell,r}q_{K,0,\ell,r},
\qquad
\Pi_hq_K=\langle q_K\rangle_hM_{n,h},
\qquad
Q_h=I-\Pi_h,
\end{equation}
where, on the uniform velocity grid,
$\omega_{\ell,r}=v_{\perp,\ell}\Delta v_\perp\Delta v_\parallel$.
The discrete neutral Maxwellian is represented only in the zeroth
gyroharmonic and satisfies
\[
\langle M_{n,h}\rangle_h=1,
\qquad
\langle v_\alpha M_{n,h}\rangle_h=0,
\qquad
\alpha\in\{x,y,z\}.
\]

In the Fourier basis, the gyroangle derivative and the collision--rotation
operator are
\begin{equation}
\label{eq:stiff-operator}
(D_{\gyro,h}q)_m=\mathrm i m q_m,
\qquad
L_h=\nuin Q_h-\Om D_{\gyro,h}.
\end{equation}
The Maxwellian projection commutes with the gyroangle derivative, giving
the exact propagator
\begin{equation}
\label{eq:projected-semigroup}
S_{\dt}=\mathrm e^{-\dt L_h}
=\Pi_h+\mathrm e^{-\nuin\dt}\mathrm e^{\Om\dt D_{\gyro,h}}Q_h.
\end{equation}
The propagator preserves the Maxwellian component while damping and
rotating the nonequilibrium component.

The temporal discretization in \cref{sec:method-update} uses the following
exponential integrator weights, related to the standard $\varphi$-functions
by $\Phi_j(z)=\varphi_{j+1}(-z)$~\cite{HochbruckOstermann2010}:
\begin{equation}
\label{eq:phis}
\begin{aligned}
\Phi_0(z)&=\frac{1-\mathrm e^{-z}}{z},
&
\Phi_1(z)&=\frac{\mathrm e^{-z}-1+z}{z^2},
\\
\Phi_2(z)&=\frac{z^2/2-z+1-\mathrm e^{-z}}{z^3}.
\end{aligned}
\end{equation}
The continuous extensions are $\Phi_0(0)=1$, $\Phi_1(0)=1/2$, and
$\Phi_2(0)=1/6$, and Taylor series are used for small $|z|$ to avoid
cancellation. With $\chi=\nuin\dt$ and
$z_m=(\nuin-\mathrm i m\Om)\dt$, the spectral decomposition of $L_h$ gives
\[
[\Phi_j(\dt L_h)q]_m=\Phi_j(z_m)q_m,
\qquad m\ne0,
\]
\[
[\Phi_j(\dt L_h)q]_0
=\Phi_j(\chi)\left(q_0-\langle q\rangle_h M_{n,h}\right)
 +\Phi_j(0)\langle q\rangle_h M_{n,h},
\qquad j=0,1,2.
\]

\subsection{Spatial and velocity-space fluxes}
\label{sec:method-interface}

Let a perpendicular spatial face $F$ separate the cells $K_L$ and $K_R$. Its
unit normal $\bn_F$ points from $K_L$ to $K_R$, and $d_F$ is the distance
between their centers. Unlimited piecewise-linear reconstruction with
centered slopes in each perpendicular direction gives the face values
$q_{F,L}$ and $q_{F,R}$ from $K_L$ and $K_R$, respectively.

The normal velocity is $v_n=\bv\cdot\bn_F$, with
$v_n^\pm=(v_n\pm|v_n|)/2$. On the gyroangle grid
$\vartheta_j=2\pi j/N_\vartheta$, where $N_\vartheta$ is even and
$N_\vartheta>2M$, define
\begin{equation}
\label{eq:angle-operators}
\begin{aligned}
(\mathcal S_Mq)_j
&=\sum_{m=-M}^{M}q_m\,\mathrm e^{\mathrm i m\vartheta_j},
\\
(\mathcal P_Mu)_m
&=\frac{1}{N_\vartheta}
  \sum_{j=0}^{N_\vartheta-1}u_j\,\mathrm e^{-\mathrm i m\vartheta_j},
\qquad |m|\le M,
\\
\mathcal V_{F,h}(q_L,q_R)
&=\mathcal P_M\!\left[
 v_n^+\mathcal S_Mq_L+v_n^-\mathcal S_Mq_R
 \right].
\end{aligned}
\end{equation}

For a discrete state $w$, $n_w=\langle w\rangle_h$ and $h_w=Q_h w$
denote its density and nonequilibrium component. In an upwind flux, the Maxwellian part contributes the term
$-\tfrac12|v_n|[n_w]_FM_{n,h}$, a numerical dissipation proportional to
the density jump across the face. This Maxwellian jump dissipation does not decrease with the
$\mathcal O(\varepsilon)$ number flux of the Hall--Pedersen scaling and can
dominate it on a fixed grid. The PH flux therefore attenuates this term,
while retaining upwinding of the nonequilibrium part, through the weight
\[
T_3(\chi)=\mathrm e^{-\chi}\left(1+\chi+\frac{\chi^2}{2}\right).
\]
It satisfies $T_3(0)=1$,
$1-T_3(\chi)=\chi^3/6+\mathcal O(\chi^4)$, and
$T_3(\chi)\to0$ as $\chi\to\infty$. The spatial PH flux is
\begin{equation}
\label{eq:ph-flux}
\mathcal F_{F,h}^{\PH}(w)
=\mathcal V_{F,h}(h_{w,L},h_{w,R})
+\mathcal P_M\!\left[
\left(v_n\{n_w\}_F-\frac12T_3(\chi)|v_n|[n_w]_F\right)M_{n,h}
\right].
\end{equation}
Here
$\{q\}_F=(q_{F,L}+q_{F,R})/2$ and $[q]_F=q_{F,R}-q_{F,L}$. The same
braces denote arithmetic averages at velocity-space faces. The corresponding
numerical number flux is
\[
\Gamma_{F,h}^{\PH}(w)
=\left\langle\mathcal V_{F,h}(h_{w,L},h_{w,R})\right\rangle_h
-T_3(\chi)d_{n,h}[n_w]_F,
\qquad
d_{n,h}=\frac12\left\langle|v_n|M_{n,h}\right\rangle_h.
\]

The density update supplements the spatial kinetic flux with a
Hall--Pedersen correction. The cell-centered Hall--Pedersen flux uses the
Maxwellian second moment
\[
\theta_h=\langle v_x^2M_{n,h}\rangle_h
        =\langle v_y^2M_{n,h}\rangle_h.
\]
The cell gradient uses reconstructed face averages, with
$(\nabla_{\perp,h}q)_i=(\{q\}_{i+1/2}-\{q\}_{i-1/2})/\Delta x$
in one dimension. This construction is applied coordinatewise in two
dimensions. In each cell, the Hall--Pedersen flux is
\begin{equation}
\label{eq:cell-hp-flux}
\begin{aligned}
\boldsymbol d_{\perp,K,h}
&=n_K\boldsymbol a_{\perp,K}-\theta_h\nabla_{\perp,h}n_K,
\\
\boldsymbol J_{\perp,K,h}
&=\frac{\nuin\boldsymbol d_{\perp,K,h}
      -\Om\bb\times\boldsymbol d_{\perp,K,h}}
      {\nuin^2+\Om^2}.
\end{aligned}
\end{equation}
The corresponding face flux is
\[
\Gamma_{F,h}^{\mathrm{cell}}
=\left\{\boldsymbol J_{\perp,h}\cdot\bn_F\right\}_F.
\]
For the leading nonequilibrium state derived in
\ref{app:discrete-hp-moment}, the kinetic number flux equals
$\Gamma_{F,h}^{\mathrm{cell}}$ by \eqref{eq:cell-face-hp-moment}.

The Pedersen and Hall drift velocities and transport coefficients are
\[
\begin{aligned}
\boldsymbol u_P
&=\frac{\nuin}{\nuin^2+\Om^2}\boldsymbol a_\perp,
&
\boldsymbol u_H
&=-\frac{\Om}{\nuin^2+\Om^2}\bb\times\boldsymbol a_\perp,
\\
D_{P,h}
&=\frac{\nuin\theta_h}{\nuin^2+\Om^2},
&
D_{H,h}
&=\frac{\Om\theta_h}{\nuin^2+\Om^2}.
\end{aligned}
\]
For constant coefficients and zero acceleration, the cell-based diffusion
has a spurious alternating null mode on even periodic grids. The compact
face flux uses two-point Pedersen diffusion to damp this mode when
$D_{P,h}>0$. In two dimensions, it is
\begin{equation}
\label{eq:compact-hp-flux-2d}
\begin{aligned}
\Gamma_{F,h}^{\mathrm{cmp}}
={}&(\boldsymbol u_{P,F}\cdot\bn_F)^+n_{F,L}
 +(\boldsymbol u_{P,F}\cdot\bn_F)^-n_{F,R}
 +(\boldsymbol u_{H,F}\cdot\bn_F)\{n\}_F
\\
&-D_{P,h}\frac{n_{K_R}-n_{K_L}}{d_F}
 +D_{H,h}\left(\bb\times\nabla_h^Fn\right)\cdot\bn_F.
\end{aligned}
\end{equation}
The Pedersen drift is upwinded, whereas the Hall drift is centered.
The face gradient uses a two-point normal difference and averaged centered
tangential differences. At an $x$-face,
\[
\begin{aligned}
(\partial_x^Fn)_{i+1/2,j}
&=\frac{n_{i+1,j}-n_{i,j}}{\Delta x},
\\
(\partial_y^Fn)_{i+1/2,j}
&=\frac{n_{i,j+1}-n_{i,j-1}
       +n_{i+1,j+1}-n_{i+1,j-1}}{4\Delta y}.
\end{aligned}
\]
The $y$-face expressions follow by interchanging $x$ and $y$.
Cancellation of the tangential differences in the finite-volume divergence gives
$\nabla_{\perp,h}\cdot(\bb\times\nabla_h^F n)=0$
for constant $\bb$ on the periodic Cartesian grid, preserving zero
divergence of the Hall pressure-gradient flux when $D_{H,h}$ is constant.
In one dimension, the upwind direction is determined by
$u_F=(\boldsymbol u_{P,F}+\boldsymbol u_{H,F})\cdot\bn_F$.
With $u_F^\pm=(u_F\pm|u_F|)/2$, the compact face flux is
\[
\Gamma_{F,h}^{\mathrm{cmp}}
=u_F^+n_{F,L}+u_F^-n_{F,R}
-D_{P,h}\frac{n_{K_R}-n_{K_L}}{d_F}.
\]
In the one-dimensional tests below, $a_y=0$, so the normal Hall drift
vanishes and this flux agrees with the one-dimensional reduction of
\cref{eq:compact-hp-flux-2d}.

In the Hall--Pedersen limit, the density-flux correction replaces the
leading cell-centered contribution by the compact face flux:
\begin{equation}
\label{eq:hp-correction}
\Gamma_{F,h}^{\mathrm{corr}}(n,\boldsymbol a)
=\left[1-T_3(\nuin\dt)\right]
 \left[
 \Gamma_{F,h}^{\mathrm{cmp}}(n,\boldsymbol a)
 -\Gamma_{F,h}^{\mathrm{cell}}(n,\boldsymbol a)
 \right].
\end{equation}
For $\nuin=0$, $T_3=1$ and $\Gamma_{F,h}^{\mathrm{corr}}=0$.

In velocity space, the radial and angular acceleration components are
$a_r=a_x\cos\gyro+a_y\sin\gyro$ and
$a_\gyro=-a_x\sin\gyro+a_y\cos\gyro$. Then
\begin{equation}
\label{eq:polar-force}
\begin{aligned}
\nabla_{\bv}\!\cdot(\boldsymbol a_\perp f)
&=\frac{1}{\vperp}\partial_{\vperp}(\vperp a_rf)
 +\frac{1}{\vperp}\partial_\gyro(a_\gyro f),
\\
(\vperp a_rf)_m
&=\frac{\vperp}{2}\left(a_-f_{m-1}+a_+f_{m+1}\right),
\qquad
(a_\gyro f)_m
=\frac{a_+f_{m+1}-a_-f_{m-1}}{2\mathrm i}.
\end{aligned}
\end{equation}
At fixed $K$ and $v_\parallel$, write $f_K=n_KM_{n,h}+h_K$ and define
the radial face flux on the gyroangle grid by
\begin{equation}
\label{eq:velocity-force-flux}
(\mathcal G_{\ell+1/2,K}^{v})_j
=a_r(\vartheta_j)\{n_K M_{n,h}\}_{\ell+1/2}
 +a_r^+(\vartheta_j)(\mathcal S_Mh_{K,\ell})_j
 +a_r^-(\vartheta_j)(\mathcal S_Mh_{K,\ell+1})_j,
\end{equation}
where $a_r^\pm=(a_r\pm|a_r|)/2$. In the collisionless Dory--Guest--Harris
tests in \cref{app:collisionless-validation}, the same radial-flux
construction centers the fixed non-Maxwellian reference $F_{0,h}$ and
upwinds $f_h-F_{0,h}$.
With
$\widehat{\mathcal G}_{\ell+1/2,K,m,r}^{v}
=(\mathcal P_M\mathcal G_{\ell+1/2,K}^{v})_{m,r}$,
the radial divergence is
\[
(D_{v_\perp,h}f)_{K,m,\ell,r}
=\frac{
 v_{\perp,\ell+1/2}\widehat{\mathcal G}_{\ell+1/2,K,m,r}^{v}
 -v_{\perp,\ell-1/2}\widehat{\mathcal G}_{\ell-1/2,K,m,r}^{v}}
 {v_{\perp,\ell}\Delta v_\perp}.
\]
The gyroangle derivative uses the modal coupling in \cref{eq:polar-force},
and the parallel acceleration flux is
\[
(\mathcal G_{r+1/2,K}^{\parallel})_{m,\ell}
=a_\parallel^+f_{K,m,\ell,r}
+a_\parallel^-f_{K,m,\ell,r+1},
\qquad
a_\parallel^\pm=\frac12(a_\parallel\pm|a_\parallel|).
\]
The radial, gyroangle, and parallel acceleration terms form the
velocity-space operator $D_{v,h}$.

For a periodic $z$ direction,
\[
\partial_{z,h}f
=\mathcal F_z^{-1}\!\left[\mathrm i k_z\mathcal F_z f\right],
\qquad
\Gamma_{\parallel,h}(w)=\langle v_\parallel w\rangle_h.
\]
The residual used in the time discretization is
\begin{equation}
\label{eq:residual}
\mathcal R_h(f,\boldsymbol a)
=-\nabla_{\perp,h}\cdot\mathcal F_h^{\PH}(f)
-v_\parallel\partial_{z,h}f-D_{v,h}(f,\boldsymbol a).
\end{equation}
The correction in \cref{eq:hp-correction} enters the numerical face flux
for the density balance, but not $\mathcal R_h$.

\subsection{Temporal discretization}
\label{sec:method-update}

\begin{figure}[!htbp]
\centering
\begin{tikzpicture}[
  x=1cm,y=1cm,>=Latex,line cap=round,line join=round,
  font=\fontsize{9}{11}\selectfont,
  pharrow/.style={->,line width=0.75pt,draw=phink!80},
  phstate/.style={line width=0.65pt,rounded corners=2pt,
    inner xsep=6pt,inner ysep=5pt,align=center},
  phbluebox/.style={phstate,draw=phblue!75,fill=phblue!8},
  phgreenbox/.style={phstate,draw=phgreen!80,fill=phgreen!9},
  phgoldbox/.style={phstate,draw=phgold!80,fill=phgold!10},
  phnotebox/.style={phstate,draw=phink!50,fill=phink!4},
  phnote/.style={align=center},
  phheading/.style={anchor=west,text=phink,font=\bfseries\fontsize{9}{11}\selectfont}
]
\definecolor{phink}{HTML}{263B50}
\definecolor{phblue}{HTML}{2575AD}
\definecolor{phgreen}{HTML}{158579}
\definecolor{phgold}{HTML}{C88B2B}
\path[use as bounding box] (0,0.00) rectangle (16.15,12.28);

\begin{scope}[yshift=2.5cm]
\node[phheading] at (0.03,9.55) {{\color{black}\fontsize{10bp}{12bp}\selectfont (a)}\quad Perpendicular spatial-flux construction};
\fill[phblue!10] (0.15,7.20) rectangle (1.25,8.55);
\fill[phgold!12] (1.25,7.20) rectangle (2.35,8.55);
\draw[phink!65,line width=0.6pt] (0.15,7.20) rectangle (2.35,8.55);
\draw[phink,line width=1pt] (1.25,7.10) -- (1.25,8.64);
\node[text=phink] at (0.70,8.28) {$K_L$};
\node[text=phink] at (1.80,8.28) {$K_R$};
\node[font=\fontsize{12}{14}\selectfont] at (0.70,7.87) {$w_{K_L}$};
\node[font=\fontsize{12}{14}\selectfont] at (1.80,7.87) {$w_{K_R}$};
\draw[pharrow] (1.25,7.47) -- (2.12,7.47);
\node at (0.82,7.47) {$\bn_F$};
\node[above=2pt,text=phink] at (1.25,8.64) {face $F$};
\node[phnote] (reconstruction) at (1.65,6.60)
  {face reconstruction\\$n_w=\langle w\rangle_h,\;h_w=Q_hw$};

\draw[pharrow,draw=phblue] (2.50,7.88) -- (3.72,7.88)
  node[midway,above=2pt,text=phink] {$\mathcal S_M$}
  node[midway,below=2pt] {$h_w$};
\begin{scope}
  \clip (4.83,7.88) circle (0.96);
  \fill[phgold!12] (3.86,6.91) rectangle (4.83,8.85);
  \fill[phblue!10] (4.83,6.91) rectangle (5.80,8.85);
\end{scope}
\draw[phink!75,line width=0.65pt] (4.83,7.88) circle (0.96);
\draw[phink!45,densely dashed,line width=0.45pt]
  (4.83,6.92) -- (4.83,8.84);
\foreach \angle in {0,22.5,45,67.5,112.5,135,157.5,180,
                     202.5,225,247.5,292.5,315,337.5}
  {\fill[phink!75] ($(4.83,7.88)+(\angle:0.96)$) circle (0.026);}
\draw[draw=phink,fill=white,line width=0.6pt] (4.83,8.84) circle (0.044);
\draw[draw=phink,fill=white,line width=0.6pt] (4.83,6.92) circle (0.044);
\node[phnote,above=2pt] at (4.83,8.84) {$v_n=0$};
\node[text=phink] at (4.35,8.19) {$h_{w,R}$};
\node[text=phink] at (5.31,8.19) {$h_{w,L}$};
\node at (4.35,7.62) {$v_n\!<\!0$};
\node at (5.31,7.62) {$v_n\!>\!0$};
\node[phnote] at (4.83,6.50) {gyroangle grid};

\node[phbluebox,minimum width=5.15cm,minimum height=1.97cm]
  (nodalflux) at (9.24,7.87)
  {$\begin{aligned}
    &v_n^+\mathcal S_Mh_{w,L}
      +v_n^-\mathcal S_Mh_{w,R}\\[3pt]
    &\quad{}+v_n\{n_w\}_FM_{n,h}\\[3pt]
    &\quad{}-\tfrac12T_3(\chi)|v_n|[n_w]_FM_{n,h}
  \end{aligned}$};
\draw[pharrow,draw=phblue] (5.94,7.88) -- (nodalflux.west);
\node[above=3pt,text=phblue!80!black] at (nodalflux.north) {nodal flux};
\draw[pharrow,draw=phgreen] (reconstruction.south) -- (1.65,5.91)
  -- (9.24,5.91) -- (nodalflux.south);
\node[above=2pt,text=phink] at (7.16,5.91)
  {$\{n_w\}_F,\ [n_w]_F$};
\node[phbluebox,minimum width=2.35cm] (distributionflux) at (14.27,7.87)
  {$\mathcal F_{F,h}^{\PH}(w)$};
\draw[pharrow,draw=phblue] (nodalflux.east) -- (distributionflux.west)
  node[midway,above=2pt,text=phink] {$\mathcal P_M$};
\node[phgreenbox,minimum width=2.35cm] (densityflux) at (14.27,6.45)
  {$\Gamma_{F,h}^{\PH}(w)$};
\draw[pharrow,draw=phgreen] (distributionflux.south) -- (densityflux.north)
  node[midway,right=3pt,text=phink] {$\langle\cdot\rangle_h$};

\draw[phink!20,line width=0.45pt] (0.05,5.63) -- (16.05,5.63);
\end{scope}

\node[phheading] at (0.03,7.70) {{\color{black}\fontsize{10bp}{12bp}\selectfont (b)}\quad Predictor and time integration};
\node[phnotebox,minimum width=1.80cm] (initial) at (1.03,6.45)
  {$f^n,\ R^0$};
\node[phbluebox,minimum width=1.20cm] (hpred) at (3.45,6.45)
  {$h^{(1)}$};
\node[phnote,above=3pt,text=phblue!80!black] at (hpred.north)
  {nonequilibrium};
\node[phgreenbox,minimum width=1.20cm] (npred) at (6.48,6.45)
  {$n^{(1)}$};
\node[phnote,below=3pt,text=phgreen!85!black] at (npred.south)
  {density};
\node[anchor=south,inner sep=1pt,text=phink] (predcorrection) at (6.48,7.04)
  {$n^n,\ \Gamma_h^{\mathrm{corr},0}$};
\draw[pharrow,draw=phgreen] (predcorrection.south) -- (npred.north);
\node[phgoldbox,minimum width=4.80cm] (predictor) at (12.00,6.45)
  {$f^{(1)}=n^{(1)}M_{n,h}+h^{(1)}$\\[2pt]
   $R^1=\mathcal R_h(f^{(1)},\boldsymbol a^1)$};
\draw[pharrow,draw=phblue] (initial.east) -- (hpred.west);
\draw[pharrow,draw=phgreen] (hpred.east) -- (npred.west)
  node[midway,above=2pt,text=phink] {\eqref{eq:phugks-update-c}};
\draw[pharrow,draw=phgold!90!black] (npred.east) -- (predictor.west);

\coordinate (inputs) at (1.95,5.25);
\draw[line width=0.75pt,phink!70] (initial.south) |- (inputs);
\draw[line width=0.75pt,phink!70] (predictor.south) |- (inputs);
\fill[phink!70] (inputs) circle (0.032);
\node[phnote,anchor=east,text=phink] at (1.72,4.40)
  {$f^n,\ R^0,\ R^1$};
\coordinate (fork) at (1.95,3.60);
\draw[line width=0.75pt,phink!70] (inputs) -- (fork);
\node[phgreenbox,minimum width=1.10cm] (average) at (3.45,3.60)
  {$\bar f$};
\node[phnote,above=3pt,text=phgreen!85!black] at (average.north)
  {time average};
\node[phbluebox,minimum width=1.10cm] (intermediate) at (3.45,2.10)
  {$\widetilde f$};
\draw[pharrow,draw=phgreen] (fork) |- (average.west);
\draw[pharrow,draw=phblue] (fork) |- (intermediate.west);

\node[phgreenbox,minimum width=3.10cm] (density) at (7.45,3.60)
  {$n^n\longmapsto n^{n+1}$\\[2pt]density update \eqref{eq:density-update}};
\draw[pharrow,draw=phgreen] (average.east) -- (density.west)
  node[midway,above=2pt,text=phink] {$\Gamma_h^{\PH}(\bar f)$};
\node[anchor=south,inner sep=1pt,text=phink] (correction) at (7.45,4.60)
  {$\tfrac12\bigl(\Gamma_h^{\mathrm{corr},0}
                    +\Gamma_h^{\mathrm{corr},1}\bigr)$};
\draw[pharrow,draw=phgreen] (correction.south) -- (density.north);

\node[phbluebox,minimum width=5.25cm] (assembled) at (12.76,2.90)
  {$f^\star=n^{n+1}M_{n,h}+Q_h\widetilde f$};
\draw[pharrow,draw=phgreen] (density.east) -| (assembled.north);
\draw[pharrow,draw=phblue] (intermediate.east) -- (9.58,2.10)
  -- (9.58,2.90) -- (assembled.west);
\node[above=2pt,text=phink] at (7.12,2.10) {$Q_h$};

\node[phgoldbox,minimum width=5.25cm] (residual) at (12.76,0.79)
  {$R^\star=\mathcal R_h(f^\star,\boldsymbol a^2)$};
\draw[pharrow,draw=phgold!90!black] (assembled.south) -- (residual.north);
\node[phbluebox,minimum width=8.15cm,minimum height=0.80cm]
  (updated) at (4.70,0.79)
  {$f^{n+1}=f^\star+\frac23\Delta t\,Q_h\Phi_1(R^\star-R^1)$};
\draw[pharrow,draw=phgold!90!black] (residual.west) -- (updated.east);
\node[phnote,below=3pt,text=phblue!80!black] at (updated.south)
  {nonequilibrium correction};
\end{tikzpicture}
\caption{Perpendicular spatial flux construction and time integration in PH-UGKS.
(a)~The nonequilibrium upwind flux, centered Maxwellian flux, and
$T_3(\chi)$-weighted Maxwellian jump term, with $\chi=\nuin\dt$, are assembled
on the gyroangle grid and projected onto the retained harmonics.
Open circles mark $v_n=0$.
(b)~The density predictor uses $h^{(1)}$ and $\Gamma_h^{\mathrm{corr},0}$
before $R^1$ is evaluated.
Green and blue paths distinguish the conservative density update from the
nonequilibrium evolution, and both enter $f^\star$.}
\label{fig:phugks-structure}
\end{figure}
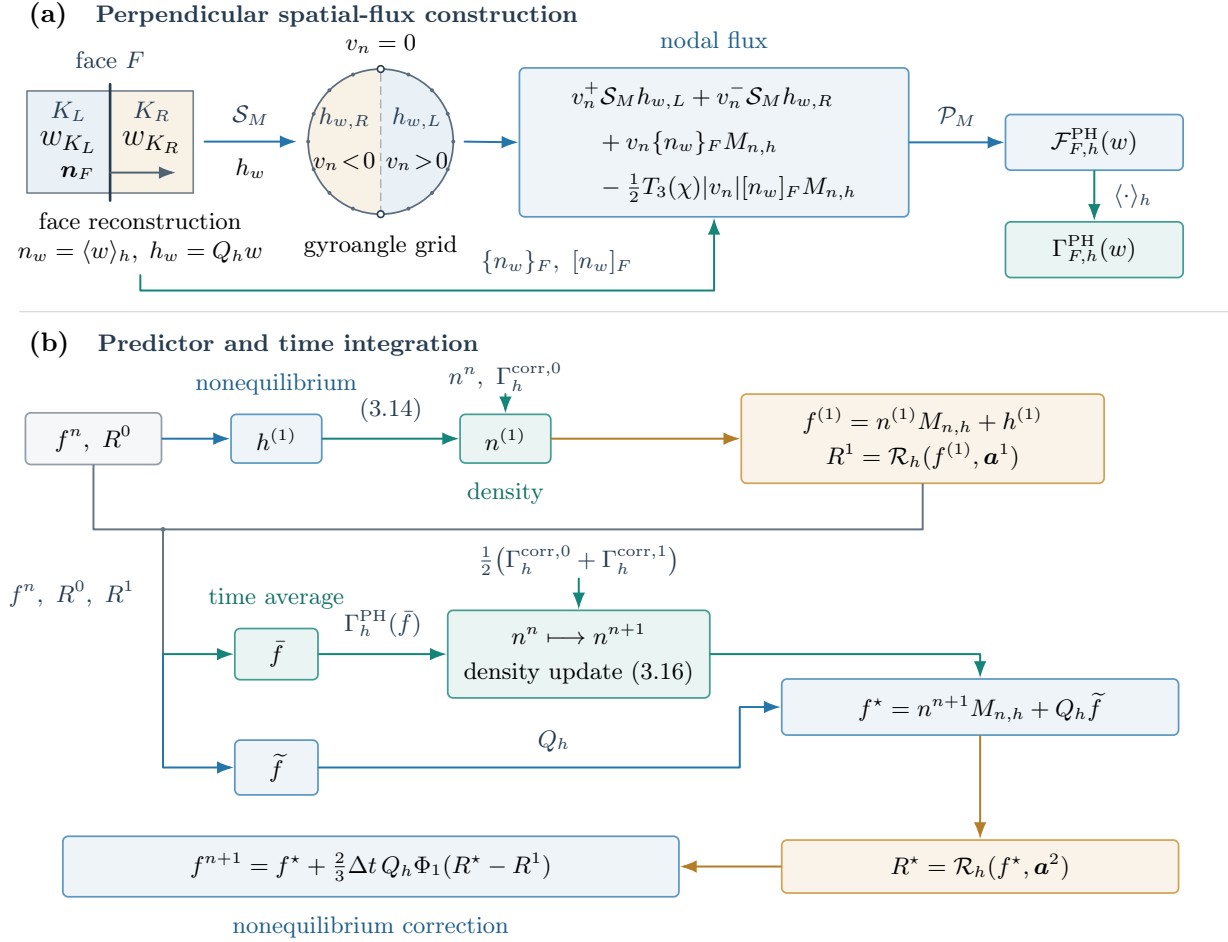

An exponential-Euler predictor for the nonequilibrium component supplies
the flux for the density predictor. The density is then advanced
conservatively with the time-averaged flux, followed by a nonequilibrium
correction at the updated density. \Cref{fig:phugks-structure} summarizes
the flux construction and these updates.

The density at $t^n$ is $n^n=\langle f^n\rangle_h$. For a self-consistent field,
$\mathcal C_h[n]$ supplies the cell and face accelerations at each stage,
with
\[
\boldsymbol a^0=\mathcal C_h[n^n],
\qquad
\boldsymbol a^1=\mathcal C_h[n^{(1)}],
\qquad
\boldsymbol a^2=\mathcal C_h[n^{n+1}].
\]
For a prescribed field,
\[
\boldsymbol a^0=\boldsymbol a(t^n),
\qquad
\boldsymbol a^1=\boldsymbol a^2=\boldsymbol a(t^{n+1}).
\]
Prescribed accelerations are evaluated at cell centers, and the compact
flux uses arithmetic averages of neighboring cell values.
Self-consistent cell accelerations in the periodic electrostatic tests are evaluated by
Fourier differentiation of the potential. For Poisson closures, the
potential is obtained by spectral inversion with its zero mode set to
zero, whereas the density-based algebraic closure is applied directly.
The self-consistent face accelerations in the compact flux are evaluated
from the discrete potential using the face gradient in \cref{sec:method-interface}.

In the stage formulas, $\Phi_j$ denotes $\Phi_j(\dt L_h)$.
The initial residual and flux correction are
$R^0=\mathcal R_h(f^n,\boldsymbol a^0)$ and
$\Gamma_h^{\mathrm{corr},0}=\Gamma_h^{\mathrm{corr}}(n^n,\boldsymbol a^0)$.
The nonequilibrium predictor is
\begin{equation}
\label{eq:phugks-update-b}
h^{(1)}=Q_h\left[S_{\dt}f^n+\dt\Phi_0R^0\right].
\end{equation}
In the Hall--Pedersen scaling, $h^{(1)}$ approaches the discrete
Chapman--Enskog state $r_h(n^n,\boldsymbol a^0)$ of
\ref{app:discrete-hp-moment}. The density predictor therefore uses
$h^{(1)}$ rather than $h^n$:
\begin{equation}
\label{eq:phugks-update-c}
n^{(1)}
=n^n-\dt\,\nabla_{\perp,h}\cdot\left[
\Gamma_h^{\PH}\!\left(n^n M_{n,h}+h^{(1)}\right)
+\Gamma_h^{\mathrm{corr},0}
\right]
-\dt\,\partial_{z,h}\Gamma_{\parallel,h}\!\left(n^n M_{n,h}+h^{(1)}\right).
\end{equation}
The residual and flux correction are then evaluated at the predictor state
$f^{(1)}=n^{(1)} M_{n,h}+h^{(1)}$:
$R^1=\mathcal R_h(f^{(1)},\boldsymbol a^1)$ and
$\Gamma_h^{\mathrm{corr},1}=\Gamma_h^{\mathrm{corr}}(n^{(1)},\boldsymbol a^1)$.

With $T_3(\nuin\dt)$ fixed over the step, linear interpolation of the
residual between $R^0$ and $R^1$ in the Duhamel formula
gives the numerical time average
\begin{equation}
\label{eq:time-averaged-distribution}
\bar f
=\Phi_0f^n+\dt(\Phi_1-\Phi_2)R^0+\dt\Phi_2R^1.
\end{equation}
The density update is
\begin{equation}
\label{eq:density-update}
n^{n+1}
=n^n-\dt\,\nabla_{\perp,h}\cdot\left[
\Gamma_h^{\PH}(\bar f)
+\frac12\left(\Gamma_h^{\mathrm{corr},0}
+\Gamma_h^{\mathrm{corr},1}\right)
\right]
-\dt\,\partial_{z,h}\Gamma_{\parallel,h}(\bar f).
\end{equation}

The conservative density update is combined with an exponential
approximation of the distribution at the new time level,
\begin{equation}
\label{eq:intermediate-distribution}
\widetilde f
=S_{\dt}f^n+\dt(\Phi_0-\Phi_1)R^0+\dt\Phi_1R^1.
\end{equation}
Replacing the Maxwellian component of $\widetilde f$ by $n^{n+1}M_{n,h}$
gives $f^\star$. The final nonequilibrium correction uses the residual at
this updated density and the corresponding stage acceleration
$\boldsymbol a^2$:
\[
f^\star=n^{n+1} M_{n,h}+Q_h\widetilde f,
\qquad
R^\star=\mathcal R_h(f^\star,\boldsymbol a^2).
\]
The distribution at $t^{n+1}$ is then
\begin{equation}
\label{eq:final-update}
f^{n+1}
=n^{n+1} M_{n,h}
+Q_h\left[
\widetilde f+\frac23\dt\Phi_1(R^\star-R^1)
\right].
\end{equation}
The final $Q_h$-projected correction leaves the density of $f^\star$
unchanged, so $\langle f^{n+1}\rangle_h=n^{n+1}$.
On periodic spatial domains, equations~\eqref{eq:density-update} and
\eqref{eq:final-update} conserve the total discrete ion number.

For the nonstiff scalar restriction $Q_h=I$, $\dot y=Ay$, replacing
$2/3$ in \cref{eq:final-update} by $c$ gives the amplification factor
$1+\xi+\xi^2/2+c\xi^3/4$, where $\xi=A\dt$.
Matching the cubic coefficient of $\mathrm e^\xi$ gives $c=2/3$.

\FloatBarrier
\section{Temporal consistency and asymptotic limits}
\label{sec:properties}

\subsection{Temporal consistency}
\label{sec:temporal-consistency}

Let $f_h(t)$ solve the semidiscrete equation
$\dot f_h=-L_hf_h+\left.\mathcal R_h(f_h,\boldsymbol a)\right|_{T_3=1}$.
Assume that the residual and the prescribed field or self-consistent field
map are twice continuously differentiable near the solution, and that the
velocity divergence has zero density moment. As $\dt\to0$ with $\nuin$
and $\Om$ held fixed, both
$T_3(\nuin\dt)-1$ and
$\Gamma_h^{\mathrm{corr}}$ are $\mathcal O(\dt^3)$, so their contributions
to one step are $\mathcal O(\dt^4)$. Starting from $f^n=f_h(t^n)$, the
predictor satisfies $f^{(1)}=f^n+\dt\dot f_h(t^n)+\mathcal O(\dt^2)$.
Thus $R^1$ approximates the reference residual at $t^{n+1}$ with error
$\mathcal O(\dt^2)$. Expansion of the weights in \cref{eq:phis} gives
\[
\bar f=f^n+\frac{\dt}{2}\dot f_h(t^n)+\mathcal O(\dt^2),
\qquad
\widetilde f=f_h(t^{n+1})+\mathcal O(\dt^3).
\]

The density update in \cref{eq:density-update} therefore satisfies
$n^{n+1}=\langle f_h(t^{n+1})\rangle_h+\mathcal O(\dt^3)$.
Hence $f^\star-f^{(1)}=\mathcal O(\dt^2)$ and
$R^\star-R^1=\mathcal O(\dt^2)$. The final correction is therefore
$\mathcal O(\dt^3)$, so
$f^{n+1}=f_h(t^{n+1})+\mathcal O(\dt^3)$.
This establishes second-order temporal consistency.

\subsection{Asymptotic limits}
\label{sec:hp-property}

In the collisionless limit $\nuin\to0$, $T_3(\nuin\dt)\to1$ and the
density-flux correction vanishes. At $\nuin=0$, the Maxwellian terms in
\cref{eq:ph-flux} combine with the nonequilibrium upwind flux to give
\[
\mathcal F_{F,h}^{\PH}(w)
=\mathcal V_{F,h}(w_{F,L},w_{F,R}).
\]
The collision--rotation propagator in \cref{eq:projected-semigroup}
reduces to
\[
(S_{\dt}q)_m=\mathrm e^{\mathrm i m\Om\dt}q_m,
\qquad |m|\le M.
\]
The propagator rotates each retained harmonic without collisional
damping, while transport and electric acceleration are incorporated
through $\mathcal R_h$.

For the Hall--Pedersen limit, we use the scaling in \cref{eq:hp-scaling} and set
$\dt=\Delta\tau/\varepsilon$, with $\beta\ge0$ and $\Delta\tau>0$.
Then $\nuin\dt=\Delta\tau/\varepsilon^2$, and the coefficients
in $\Gamma_{F,h}^{\mathrm{cmp}}$ are proportional to $\varepsilon$. We write
$\widetilde\Gamma_{F,h}^{\mathrm{cmp}}
=\varepsilon^{-1}\Gamma_{F,h}^{\mathrm{cmp}}$.
For smooth $n$ and $\boldsymbol a$, with face accelerations accurate to
second order,
this flux approximates the Hall--Pedersen flux in \eqref{eq:hp-density-limit}
to second order in space, with $\theta_n$ replaced by $\theta_h$.

For a discrete gyroharmonic state $q$, let
\[
\|q\|_h
=
\max_K\left(
2\pi\sum_{m=-M}^{M}\sum_{\ell,r}
\omega_{\ell,r}|q_{K,m,\ell,r}|^2
\right)^{1/2}.
\]

\Needspace{6\baselineskip}
The one-step and finite-step results below are proved in
\ref{app:proof-hp-theorem}.

\begin{theorem}[Hall--Pedersen limit of the density update]
\label{prop:HP}
Consider the perpendicular dynamics on a one- or two-dimensional periodic
uniform Cartesian grid with the discretization of \cref{sec:scheme}. Let the
complete velocity operator satisfy the conditions stated in
\ref{app:discrete-hp-moment}. Assume that $\mathcal R_h$ is Lipschitz on
fixed bounded sets containing all stage arguments $(f,\boldsymbol a)$
and their Maxwellian counterparts $(\Pi_h f,\boldsymbol a)$.
In the self-consistent case, assume that $\mathcal C_h$ is Lipschitz on
the corresponding bounded density sets. The Lipschitz constants are
independent of $\varepsilon$ for $0<\varepsilon\le\varepsilon_0$.
For a prescribed acceleration, let
$\boldsymbol a^\varepsilon(t)=\overline{\boldsymbol a}(\varepsilon t)$,
where $\overline{\boldsymbol a}$ is locally Lipschitz. If
\[
f^n=n^n M_{n,h}+h^n,
\qquad
\|n^n\|_\infty+\varepsilon^{-1}\|h^n\|_h\le C,
\]
with $C$ independent of $\varepsilon$, then the predictor satisfies
\begin{equation}
\label{eq:HP-predictor}
n^{(1)}
=
n^n-\Delta\tau\,\nabla_{\perp,h}\cdot
\widetilde\Gamma_h^{\mathrm{cmp}}(n^n,\boldsymbol a^0)
+\mathcal O(\varepsilon),
\end{equation}
and the density after one time step satisfies
\begin{equation}
\label{eq:HP-estimate}
n^{n+1}
=
n^n-\frac{\Delta\tau}{2}\nabla_{\perp,h}\cdot\left[
\widetilde\Gamma_h^{\mathrm{cmp}}(n^n,\boldsymbol a^0)
+\widetilde\Gamma_h^{\mathrm{cmp}}(n^{(1)},\boldsymbol a^1)
\right]+\mathcal O(\varepsilon).
\end{equation}
The $\mathcal O(\varepsilon)$ remainders are uniform over the grid, and
\[
\|Q_h f^{n+1}\|_h=\mathcal O(\varepsilon).
\]
Hence the limiting density update is the Heun discretization with the compact
Hall--Pedersen flux defined in \cref{sec:method-interface}.
\end{theorem}

In the Hall--Pedersen scaling the $T_3$-weighted Maxwellian jump vanishes,
\cref{eq:hp-correction} replaces the leading cell-centered flux
$\Gamma_{F,h}^{\mathrm{cell}}$ of \ref{app:discrete-hp-moment} by
$\Gamma_{F,h}^{\mathrm{cmp}}$, and the limiting time-average weights yield
the arithmetic mean of the two stage fluxes.

\Needspace{4\baselineskip}
Let $f_{\varepsilon,h}^j$ denote the PH-UGKS state at slow time
$\tau^j=j\Delta\tau$, and let
$n_{\varepsilon,h}^j=\langle f_{\varepsilon,h}^j\rangle_h$.
Write $N_h^j$ for the solution of the compact Hall--Pedersen--Heun
scheme on the same grid, with the same $\Delta\tau$ and field
discretization. Assume the initial bound in \Cref{prop:HP} and the
same uniform Lipschitz estimates on fixed bounded sets containing all
kinetic and limiting stages. For $N$ fixed independently of $\varepsilon$,
iterating the one-step estimates gives
\[
\begin{aligned}
\max_{0\le j\le N}\|n_{\varepsilon,h}^j-N_h^j\|_\infty
&\le C_N\left(\|n_{\varepsilon,h}^0-N_h^0\|_\infty+\varepsilon\right),
\\
\max_{0\le j\le N}\|Q_hf_{\varepsilon,h}^j\|_h
&\le C_N\varepsilon,
\end{aligned}
\]
with $C_N$ independent of $\varepsilon$.

These estimates establish asymptotic preservation of the Hall--Pedersen
density limit under the stated assumptions.

\section{Numerical results}
\label{sec:results}

The tests assess spatial and temporal accuracy, finite-Larmor-radius kinetic
response, the Hall--Pedersen density limit, and nonlinear self-consistent
coupling. \Cref{app:collisionless-validation} presents comparisons with an
independently implemented rotating-grid solver and tests of the
Dory--Guest--Harris instability.

All tests are uniform along the magnetic field with $E_\parallel=0$.
The common Maxwellian dependence of the initial distribution and BGK
equilibrium on $v_\parallel$ permits reduction to
$f(\bx_\perp,\vperp,\gyro,t)$. Here $M_n$ denotes the perpendicular
Maxwellian with $\theta_n=1$, and $M_h$ consists of its exact radial-cell
averages with respect to $\vperp\,\mathrm d\vperp$. The discrete second moment
$\theta_h$ is evaluated from these averages and the radial cell centers.
For a gyroharmonic coefficient array $g_{r,m}$, the velocity-space norm is
\begin{equation*}
\|g\|_{2,v}^2=2\pi\sum_r w_r\sum_{m=-M}^{M}|g_{r,m}|^2.
\end{equation*}
Here $r$ indexes the radial cells of the reduced perpendicular problem,
and $w_r$ is the corresponding quadrature weight.
The gyroharmonic weights are
\begin{equation}
e_{0,r}=w_r|g_{r,0}|^2,\qquad
e_{|m|,r}=w_r\left(|g_{r,m}|^2+|g_{r,-m}|^2\right),\qquad
\mathcal E_{|m|}=2\pi\sum_r e_{|m|,r},
\label{eq:paired-content}
\end{equation}
where $\mathcal E_{|m|}$ denotes the contribution at gyroharmonic order
$|m|$ to the squared velocity norm $\|g\|_{2,v}^{2}$.

\subsection{Spatial and temporal accuracy}
\label{sec:res-accuracy}

To assess spatial and temporal accuracy, we solve the perpendicular form of \cref{eq:model}
on the periodic domain $[0,4\pi]^2$ with a prescribed uniform acceleration.
The initial state is $f_{0,h}=n_{0,h}M_h$,
where $n_{0,h}=1+0.05p_h$ and $p_h$ is a band-limited density perturbation.
Its Fourier support is
\begin{equation*}
\mathcal K_J=\{(p,q)\in\mathbb Z^2:0<p^2+q^2\leq J^2\},
\qquad J\in\{2,6\},
\qquad \bm k_{pq}=\tfrac12(p,q),
\end{equation*}
with the elliptic Gaussian coefficients
\begin{equation}
\widehat p_{pq}=\exp\!\left[-\frac12
\left(1.5^2(k'_{pq,1})^2+0.9^2(k'_{pq,2})^2\right)\right]
\mathrm e^{-\mathrm i\bm k_{pq}\cdot\bm x_c},
\quad
\bm k'_{pq}=R(-\pi/6)\bm k_{pq},
\quad
\bm x_c=(2\pi,2\pi).
\label{eq:ap-packet}
\end{equation}
Here $R(\alpha)$ denotes planar rotation through angle $\alpha$.
For each support $\mathcal K_J$, the real inverse Fourier sum is normalized
separately to unit $L^\infty$ amplitude before exact spatial cell averaging,
giving $p_h$. The accuracy tests use $J=2$, and the initial density in
\cref{sec:res-ap-packet} uses the same Fourier construction with $J=6$.

The velocity grid is $(N_{\vperp},M,N_\gyro)=(32,8,32)$ with
$\vperp^{\max}=8$. In both tests, the full-distribution error is normalized
by the norm of the spatially varying part of the reference distribution
at the comparison time:
$E_f=\|f_h-f_{\mathrm{ref},h}\|_{2,h}/
\|f_{\mathrm{ref},h}-\langle f_{\mathrm{ref},h}\rangle_{\bm x}\|_{2,h}$,
where $\langle\cdot\rangle_{\bm x}$ denotes the spatial average and
$\|g\|_{2,h}^2=\Delta x\Delta y\sum_K\|g_K\|_{2,v}^2$.

\begin{figure}[!htbp]
  \centering
  \includegraphics[width=0.88200\textwidth]{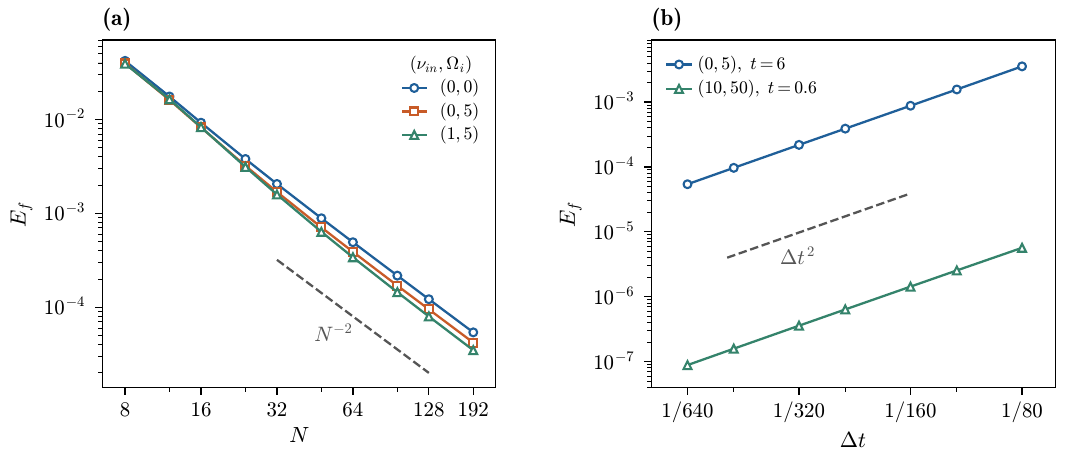}
  \caption{Full-distribution accuracy. (a)~Error versus $N$ at
  $t_f=1/2$, with $N_x=N_y=N$ and $\dt=1/(8N)$.
  The reference is the exact solution \eqref{eq:joint-mode-reference}
  of the velocity-discretized equation.
  (b)~Error versus $\dt$ at fixed spatial and velocity resolution,
  for the indicated collision and gyrofrequencies $(\nuin,\Om)$ and
  observation times. The reference is a same-grid PH-UGKS solution with
  $\dt_{\mathrm{ref}}=1/5120$.
  Dashed segments indicate $N^{-2}$ and $\dt^2$.}
  \label{fig:accuracy-refinement}
\end{figure}

Joint grid--time refinement uses three parameter sets:
$(\nuin,\Om)=(0,0)$ with $\bm a=\bm0$, and $(0,5)$ and $(1,5)$ with
$\bm a=\bm a_0=(2\bm e_x+\bm e_y)/\sqrt5$.
The first case is free transport; the other two include electric
acceleration and cyclotron rotation, with BGK relaxation in the third.
For the prescribed constant acceleration, the velocity-discretized equation
is linear with constant coefficients, so its spatial Fourier modes satisfy
\begin{equation}
\widehat f_{\bm k}(t)
=\exp(tA_{v,\bm k})\widehat f_{\bm k}(0),\qquad
A_{v,\bm k}=-L_h-\mathrm i(k_xV_{x,h}+k_yV_{y,h})-D_{v,h}.
\label{eq:joint-mode-reference}
\end{equation}
Here $V_{x,h}$ and $V_{y,h}$ represent multiplication by $v_x$ and $v_y$
in the retained gyroharmonics, and $D_{v,h}$ is the discrete force operator
at the prescribed acceleration. The matrix exponentials are evaluated in
double precision using a scaling-and-squaring algorithm with Pad\'e
approximation. Every spatial mode, including the mean, is evolved;
exact spatial cell averaging gives $f_{\mathrm{ref},h}$.
With the velocity discretization held fixed, this test measures the joint
spatial and temporal convergence of the distribution function against the
exact solution of the velocity-discretized equation. We use
$N_x=N_y=N\in\{8,12,16,24,32,48,64,96,128,192\}$ with
$\dt=1/(8N)$ and evaluate the error at $t_f=1/2$.
The three sequences give fitted rates of
$2.02$, $2.04$, and $2.09$, respectively (\cref{fig:accuracy-refinement}(a)).

For the temporal accuracy test, we fix $N_x=N_y=32$ and consider
$(\nuin,\Om)=(0,5)$ and $(10,50)$, both with $\bm a=\bm a_0$.
The seven time steps are $1/80$, $1/120$, $1/160$,
$1/240$, $1/320$, $1/480$, and $1/640$.
At each observation time, $f_{\mathrm{ref},h}$ is obtained with the same scheme
on the same spatial--velocity grid at $\dt_{\mathrm{ref}}=1/5120$.
The observed rate is $2.01$ for the collisionless case at $t=6$ and
for the collisional case at $t=0.6$ (\cref{fig:accuracy-refinement}(b)),
with $\Om t=30$ in both cases.
The collisional histories also give $2.01$ at $t=0.1$ and $0.3$,
and $2.96$ at $t=1.25$.
Setting $T_3=1$ in both the numerical and reference solutions gives
second-order convergence at $t=1.25$.
The observed second-order convergence is consistent with the analysis in
\cref{sec:temporal-consistency}.

\subsection{Ion Bernstein waves}
\label{sec:res-collisionless}
\label{sec:res-kend}

The one-dimensional electrostatic problem uses a quasineutral closure with
a linearized adiabatic-electron response~\cite{Schild2025},
\begin{equation}
\partial_t f+v_x\partial_x f+E_x\partial_{v_x}f-\Om\partial_\gyro f=0,
\qquad
E_x=-\partial_x\phi,
\qquad
\phi=\delta n,
\label{eq:nibw-model}
\end{equation}
with $\nuin=0$ and $\Om=\rho_i=v_{\mathrm{th}}=1$.  Equal-amplitude density
perturbations are assigned to every positive spatial Fourier mode below the
Nyquist wavenumber,
\begin{equation}
f(x,\vperp,\gyro,0)=M_n(\vperp)
\left[1+10^{-6}\sum_{q=1}^{N_x/2-1}\cos(k_q x)\right],
\qquad k_q=\frac{2\pi q}{L}.
\label{eq:nibw-seed}
\end{equation}
The domain has length $L=20\pi$.  We use
$(N_x,N_{\vperp},M,N_\gyro)=(128,32,16,64)$,
$\vperp^{\max}=6$, $\dt=0.02$, and $T=80$.

The continuum frequencies are the positive roots of the ion Bernstein
dispersion relation~\cite{Bernstein1958},
\begin{equation}
1-\sum_{p\geq1}\frac{2p^2\Gamma_p(b)}{\omega^2-p^2}=0,
\qquad
\Gamma_p(b)=\mathrm e^{-b}I_p(b),
\qquad b=(k_\perp\rho_i)^2.
\label{eq:nibw-dispersion}
\end{equation}
For $b>0$, each interval $\ell<\omega<\ell+1$ contains one continuum root.
An order-10 matrix-pencil fit over $0\leq t\leq80$ is applied to the
potential history at each resolved wavenumber~\cite{HuaSarkar1990}.
For $\ell=1,\ldots,4$, branch $\ell$ is identified with the strongest
positive-frequency pole in $\ell<\omega<\ell+1$.

\begin{figure}[p]
  \centering
  \includegraphics[width=0.88200\textwidth,trim=0bp 6bp 0bp 0bp,clip]{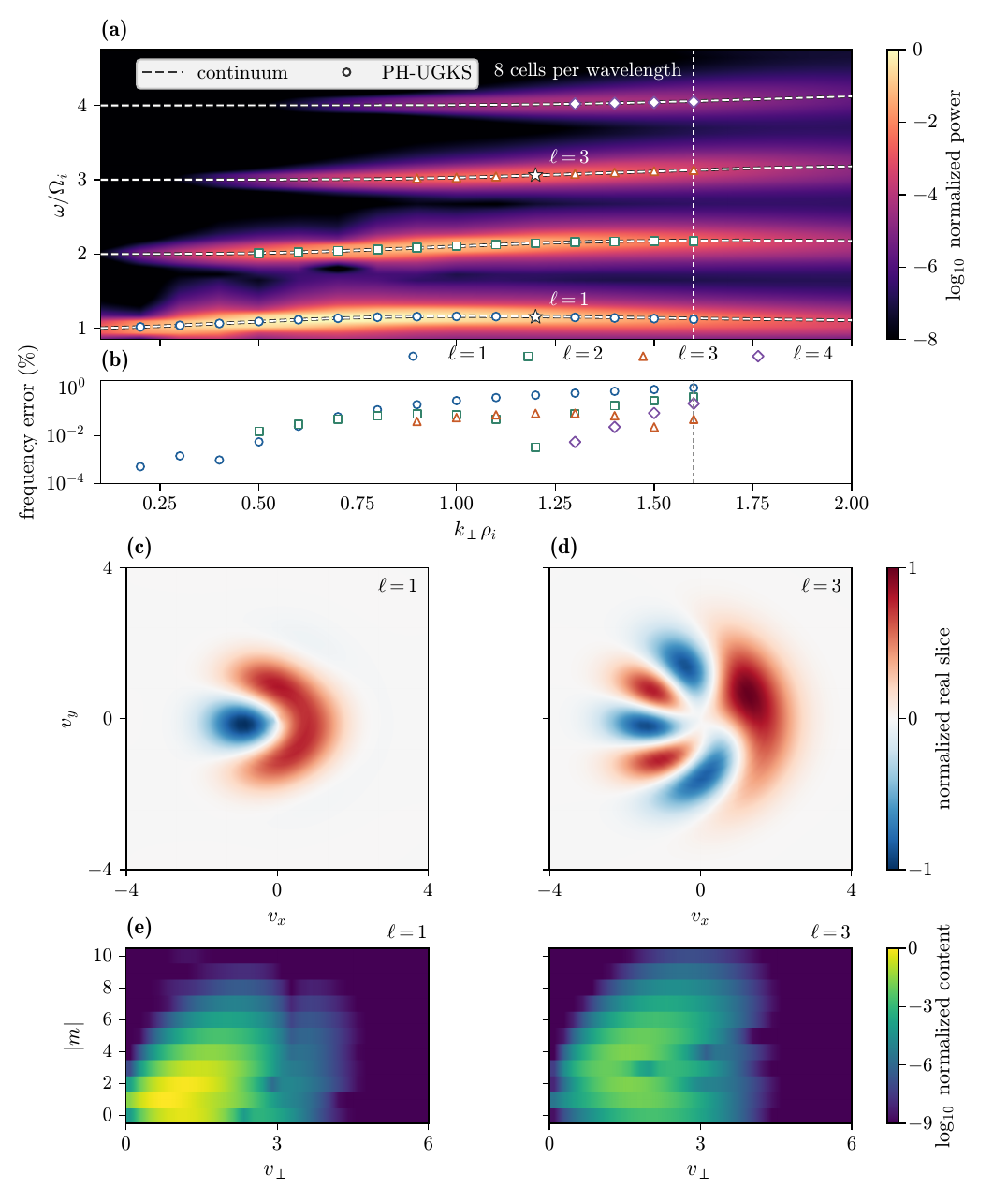}
  \caption{Collisionless ion Bernstein spectrum and velocity-space structure.
  (a)~Hann-windowed potential power normalized by its global maximum,
  with continuum frequencies (dashed lines) and numerical
  fits (symbols). The vertical guide marks eight cells per wavelength, and
  stars identify branches 1 and 3 at $k_\perp\rho_i=1.2$.
  (b)~Relative frequency errors for retained pairs.
  (c,d)~Real reconstructions from \cref{eq:nibw-demod}, each normalized by
  its full-domain maximum absolute value.
  (e)~Paired gyroharmonic content $e_{|m|,r}$ for $\ell=1$ (left) and $\ell=3$ (right),
  normalized by their common maximum. Only $|m|\leq10$ of the retained
  $|m|\leq16$ modes are shown.}
  \label{fig:nibw-final}
\end{figure}

\begin{samepage}
For the branch-resolved velocity-space states, Hann-weighted demodulation at
the fitted frequency gives
\begin{equation}
\widehat f_m^{\,B}(\vperp)=
\frac{\displaystyle\int_0^T\varpi_H(t)\widehat f_{q,m}(\vperp,t)
\mathrm e^{\mathrm i\omega_{q\ell}^{h}t}\,\mathrm dt}
{|\widehat\phi_q(0)|\displaystyle\int_0^T\varpi_H(t)\,\mathrm dt},
\qquad
\varpi_H(t)=\frac12\left[1-\cos\left(\frac{2\pi t}{T}\right)\right],
\label{eq:nibw-demod}
\end{equation}
where $\widehat f_{q,m}$ is the spatial Fourier coefficient of the full
distribution.  The two states shown correspond to $k_\perp\rho_i=1.2$, with
$\ell=1$ and $3$.  The maps show
$2\operatorname{Re}\sum_m\widehat f_m^{B}(\vperp)
\mathrm e^{\mathrm im\gyro}$ at the same demodulation phase, and the
paired gyroharmonic content in \cref{eq:paired-content} is evaluated with
$g_m=\widehat f_m^B$.
\end{samepage}

The maximum relative frequency difference over the 39 branch--wavenumber
pairs with at least eight cells per wavelength and a fitted amplitude
above $10^{-2}|\widehat\phi_q(0)|$ is $0.998\%$ (\cref{fig:nibw-final}).
At the same wavenumber, the branch-3 state has
a larger relative contribution from higher gyroharmonics than branch 1
(\cref{fig:nibw-final}(e)),
illustrating velocity-space structure beyond the flux-carrying
$m=\pm1$ pair.

\subsection{Finite-collisionality dynamics}
\label{sec:res-collisional}

The driven response is examined along two parameter paths: varying
collisionality at fixed $k_\perp\rho_i$, and increasing the collision and
gyrofrequencies at a fixed ratio while holding the absolute wavenumber
and drive frequency constant. The baseline time steps resolve spatial
transport and the drive, whereas the large-step tests retain drive
resolution while allowing $\vperp^{\max}\dt/\Delta x>1$.

\subsubsection{Driven response at fixed \texorpdfstring{$k_\perp\rho_i$}{k⊥ρᵢ}}
\label{sec:res-collision}

On one periodic wavelength, the distribution satisfies the driven
Vlasov--BGK equation
\begin{equation}
\partial_t f+v_x\partial_x f+a_x(x,t)\partial_{v_x}f
-\Om\partial_\gyro f=\nuin(nM_n-f),
\qquad a_x=-\partial_x\phi_{\mathrm{ext}}.
\label{eq:patha-model}
\end{equation}
The prescribed potential is
\begin{equation}
\phi_{\mathrm{ext}}(x,t)=2\times10^{-3}s(t)
\sin(\widehat\omega t)\cos(kx),
\qquad k\rho_i=\sqrt2,\qquad \widehat\omega=1.1,\qquad \Om=1,
\label{eq:patha-drive}
\end{equation}
where $T_d=2\pi/\widehat\omega$ and
\begin{equation}
s(t)=
\begin{cases}
10\xi^3-15\xi^4+6\xi^5,
&0\leq t\leq2T_d,\quad \xi=t/(2T_d),\\
1,&t>2T_d.
\end{cases}
\label{eq:patha-ramp}
\end{equation}
The initial state is $M_n$.  The thirteen collision ratios
\begin{equation*}
\eta=\frac{\nuin}{\Om}\in
\{0.03,0.05,0.1,0.2,0.3,0.5,0.8,1,1.5,2,3,5,8\}
\end{equation*}
use $(N_x,N_{\vperp},M,N_\gyro)=(24,48,16,64)$,
$\vperp^{\max}=8$, $\dt=T_d/320$, and $T=39T_d$.

\begin{figure}[!htbp]
  \centering
  \includegraphics[width=0.88200\textwidth]{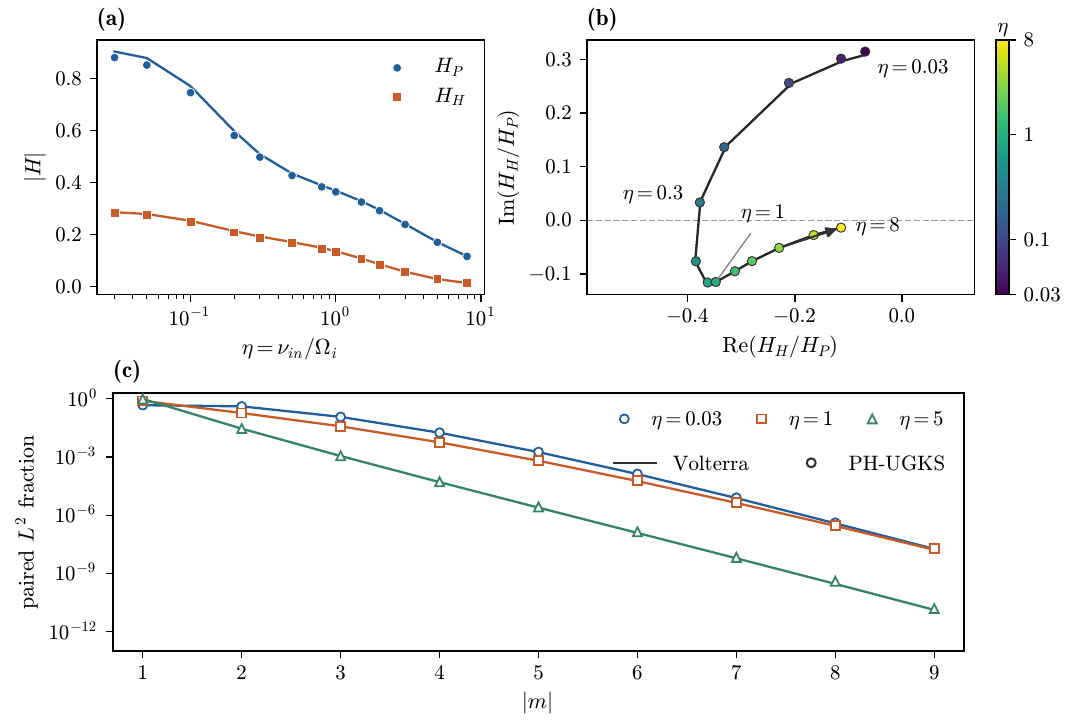}
  \caption{Finite-collisionality response at
  $k_\perp\rho_i=\sqrt2$. (a)~Magnitudes of the drive-aligned and
  transverse transfer coefficients. (b)~$H_H/H_P$ in the complex plane,
  with a logarithmic $\eta$ color scale and an arrow indicating
  increasing $\eta$. (c)~Paired gyroharmonic
  spectra at $\eta=0.03$, $1$, and $5$.
  The Volterra coefficients in (c) are averaged over the numerical radial cells.
  Each spectrum is normalized by its own nongyrotropic content
  $\sum_{m=1}^{M}\mathcal E_{|m|}$, with $M=16$, and only
  $1\leq|m|\leq9$ is shown.
  Solid lines denote the characteristic--Volterra reference and symbols denote
  PH-UGKS in all panels.}
  \label{fig:patha-summary}
\end{figure}

The characteristic--Volterra reference is obtained from the Vlasov--BGK
equation linearized about $M_n$, using rotating characteristics.
For the driven spatial mode, the Duhamel formula gives
\begin{equation}
f_k(\vperp,\gyro,t)=\int_0^t\mathrm e^{-\nuin\tau}
\mathrm e^{-\mathrm i\Phi(\vperp,\gyro,\tau)}
\left[\nuin n_k(t-\tau)M_n
+a_k(t-\tau)\frac{v_x(\gyro+\Om\tau)}{\theta_n}M_n\right]\mathrm d\tau,
\label{eq:volterra-characteristic}
\end{equation}
where
\begin{equation*}
\Phi(\vperp,\gyro,\tau)=\frac{k\vperp}{\Om}
\left[\sin(\gyro+\Om\tau)-\sin\gyro\right].
\end{equation*}
Its density moment closes through the scalar Volterra equation
\begin{equation}
n_k(t)=\nuin\int_0^t\mathrm e^{-\nuin\tau}K_n(\tau)n_k(t-\tau)\,\mathrm d\tau
+\int_0^t\mathrm e^{-\nuin\tau}K_a(\tau)a_k(t-\tau)\,\mathrm d\tau,
\label{eq:volterra-density}
\end{equation}
with
\begin{equation*}
K_n(\tau)=\mathrm e^{-k^2\rho_i^2(1-\cos\Om\tau)},
\qquad
K_a(\tau)=-\mathrm i\frac{k}{\Om}\sin(\Om\tau)K_n(\tau).
\end{equation*}

\begin{figure}[!htbp]
  \centering
  \includegraphics[width=0.80539\textwidth]{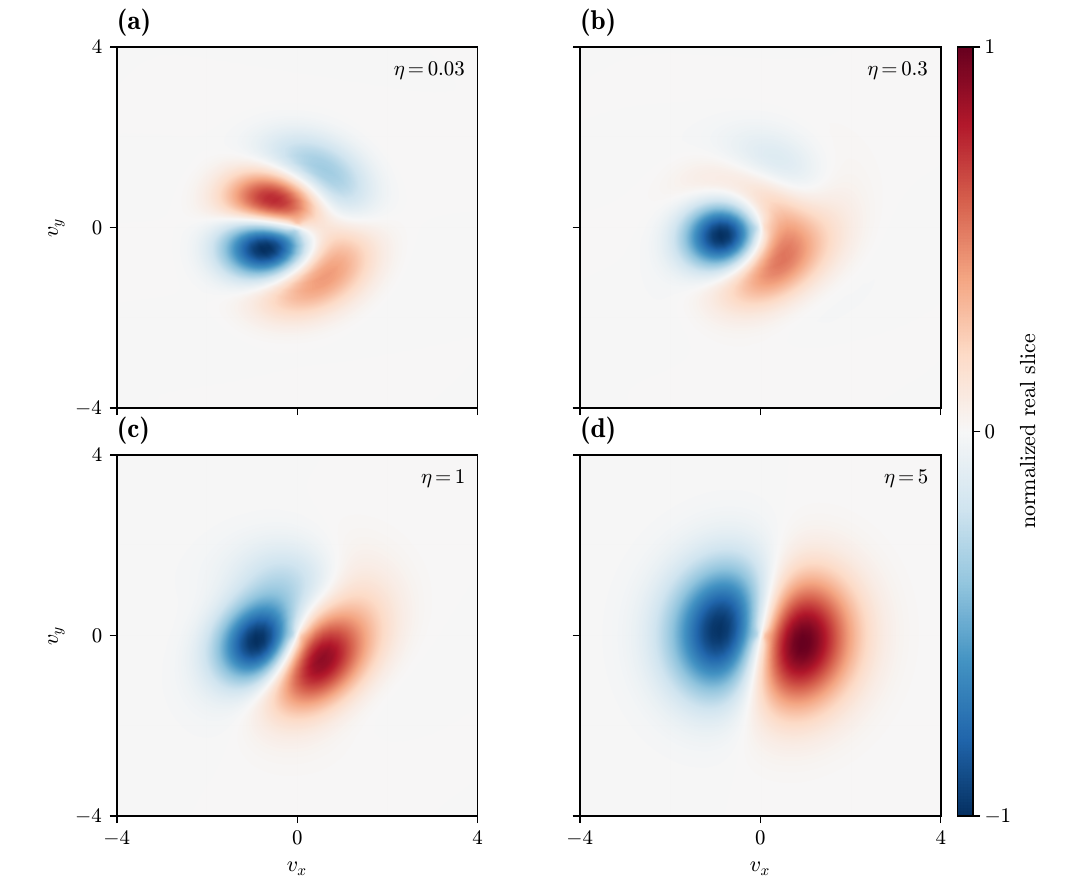}
  \caption{Nongyrotropic drive-frequency structure from PH-UGKS at
  $\eta=0.03$, $0.3$, $1$, and $5$. The maps show
  $2\operatorname{Re}\sum_{m\ne0}\widehat{\delta f}_m^{\,\omega}
  \mathrm e^{\mathrm im\gyro}$ at a common drive phase.
  Each reconstruction is normalized by its maximum absolute value over
  the full computational velocity domain.}
  \label{fig:patha-states}
\end{figure}

The perpendicular ion flux and gyroharmonic coefficients are obtained
from this linear-response solution.

Let $\widehat a_k$ and $\widehat\Gamma_{j,k}$ be the driven Fourier
coefficients, with $j=P$ aligned with the drive and
$j=H$ along $\bb\times\bm e_P$.  Over $W_1=[31T_d,35T_d]$, define
\begin{equation}
H_j=\frac{\displaystyle\int_{W_1}\widehat\Gamma_{j,k}(t)
\mathrm e^{\mathrm i\widehat\omega t}\,\mathrm dt}
{\displaystyle\int_{W_1}\widehat a_k(t)
\mathrm e^{\mathrm i\widehat\omega t}\,\mathrm dt},
\label{eq:patha-transfer}
\end{equation}
where $\bm H=(H_P,H_H)$.  The relative transfer difference is
$E_V=\|\bm H_{\PH}-\bm H_V\|_2/\|\bm H_V\|_2$, with the subscript $V$
denoting the Volterra reference.
Replacing $\widehat\Gamma_{j,k}$ in \cref{eq:patha-transfer} by
$\widehat{\delta f}_{k,m}$, with $\delta f=f-M_h$, defines
$\widehat{\delta f}_m^{\,\omega}$. Its $m\ne0$ components are used for the
spectra and velocity maps.

Both PH-UGKS and the Volterra reference show $H_H/H_P$ crossing the
negative real axis between $\eta=0.3$ and $0.5$, corresponding to
antiphase drive-frequency flux components at the crossing
(\cref{fig:patha-summary}(b)). In contrast, the instantaneous
Hall--Pedersen relation gives the real ratio $H_H/H_P=-1/\eta$ for every $\eta$.

In both solutions, the $m=\pm1$ pair carries approximately $46\%$,
$77\%$, and $97\%$ of the nongyrotropic content at $\eta=0.03$, $1$, and $5$,
respectively (\cref{fig:patha-summary}(c)).
\Cref{fig:patha-states} shows the corresponding transition from
multi-lobed to predominantly dipolar velocity-space structure.

On the baseline grid, the maximum relative transfer difference from the
Volterra reference is $E_V=4.19\%$ at $\eta=0.05$.
At $\eta=0.03$, joint spatial and temporal refinement to $N_x=96$ and
$\dt=T_d/1280$ reduces $E_V$ from $4.14\%$ to $1.11\%$, with the velocity
grid unchanged. Further radial refinement from $N_{\vperp}=48$ to $96$,
at the same spatial resolution and time step, reduces $E_V$ to $0.37\%$.
Together with the phase relation and gyroharmonic spectra in
\cref{fig:patha-summary}, these comparisons support the collision-dependent
finite-frequency kinetic response beyond the instantaneous Hall--Pedersen
relation.

\FloatBarrier
\subsubsection{Finite-frequency approach to the Hall--Pedersen limit}
\label{sec:res-pathb}

For the second parameter path, the collision and gyrofrequencies satisfy
\begin{equation*}
\nuin=\Om=\Lambda,
\qquad
\Lambda\in\{1,3,10,30,100,300,1000\}.
\end{equation*}

\begin{figure}[!htbp]
  \centering
  \includegraphics[width=0.88200\textwidth]{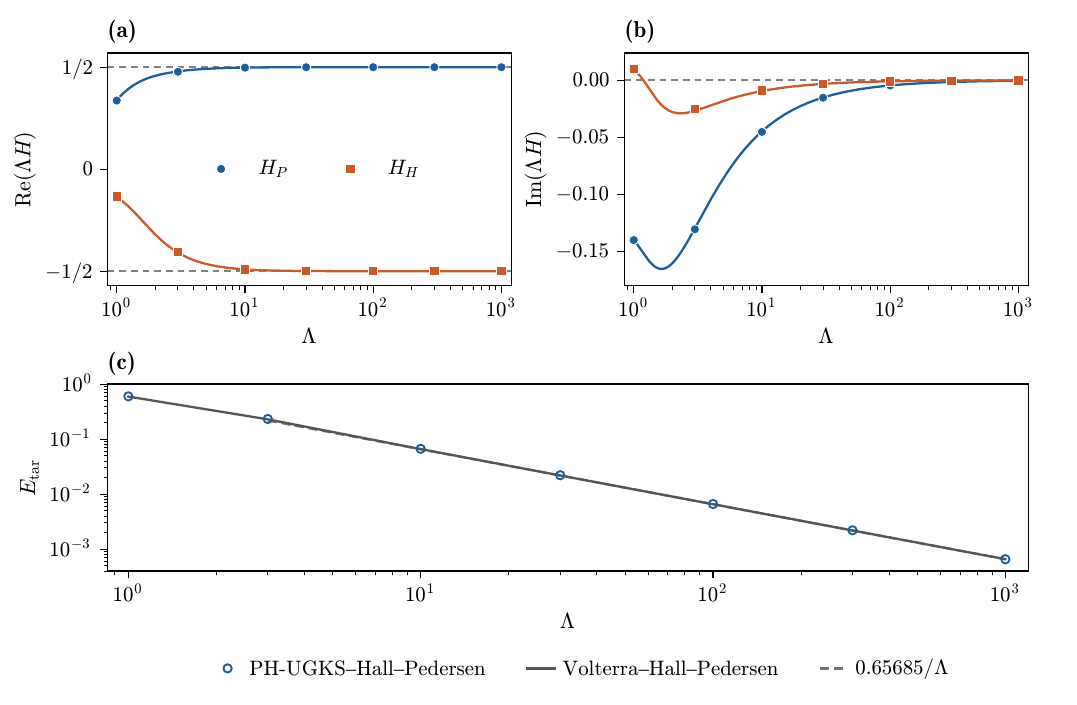}
  \caption{Finite-frequency approach to the Hall--Pedersen coefficients at
  $\nuin=\Om=\Lambda$, with $k=\sqrt2$ and $\widehat\omega=1.1$ fixed.
  (a)~Real parts of $\Lambda H_P$ and $\Lambda H_H$.
  (b)~Their imaginary parts. In (a,b), solid lines are the
  characteristic--Volterra reference, symbols are PH-UGKS, and horizontal
  dashed lines mark the Hall--Pedersen values.
  (c)~Distances $E_{\mathrm{tar}}^{\PH}$ and $E_{\mathrm{tar}}^V$ to the
  Hall--Pedersen values. The dashed line is the leading term in
  \cref{eq:pathb-asymptotic}.}
  \label{fig:pathb-response}
\end{figure}

The remaining parameters are those of \cref{sec:res-collision}, with the
absolute wavenumber $k=\sqrt2$ and drive frequency $\widehat\omega=1.1$
held fixed. The baseline time step is $\dt=T_d/1024$. Thus $\nuin/\Om=1$ is fixed, whereas
$k_\perp\rho_i$ and $\widehat\omega/\Om$ decrease in proportion to
$\Lambda^{-1}$.
The integrals defining the numerical transfer coefficients are evaluated
by the composite trapezoidal rule.

At this ratio the limiting value of $\Lambda\bm H$ is
$\bm H_{HP}=(1/2,-1/2)$.  For PH-UGKS and the Volterra reference,
the normalized distance to this target is
$E_{\mathrm{tar}}^s=\|\Lambda\bm H_s-\bm H_{HP}\|_2/\|\bm H_{HP}\|_2$,
$s\in\{\PH,V\}$.
The large-$\Lambda$ expansion of the linear-response reference gives
\begin{equation}
E_{\mathrm{tar}}^V
=\frac{\sqrt{\widehat\omega^{\,4}
 +(\widehat\omega^{\,2}-\theta_n k^2)^2}}
 {2\widehat\omega\Lambda}
+\Ord(\Lambda^{-2}).
\label{eq:pathb-asymptotic}
\end{equation}
For $k=\sqrt2$, $\widehat\omega=1.1$, and $\theta_n=1$, the leading term
is $0.65685/\Lambda$. The leading departure from the Hall--Pedersen
coefficients arises from ion inertia and the pressure gradient associated
with the density perturbation.

PH-UGKS follows the finite-frequency Volterra response with a maximum
relative transfer difference $E_V$ of $1.49\%$. The normalized distance
$E_{\mathrm{tar}}^{\PH}$ decreases from $0.602$ at $\Lambda=1$ to
$6.55\times10^{-4}$ at $\Lambda=1000$, following the $\Lambda^{-1}$
trend of the reference (\cref{fig:pathb-response}).

\begin{figure}[!htbp]
  \centering
  \includegraphics[width=0.88200\textwidth,trim=0bp 18bp 0bp 0bp,clip]{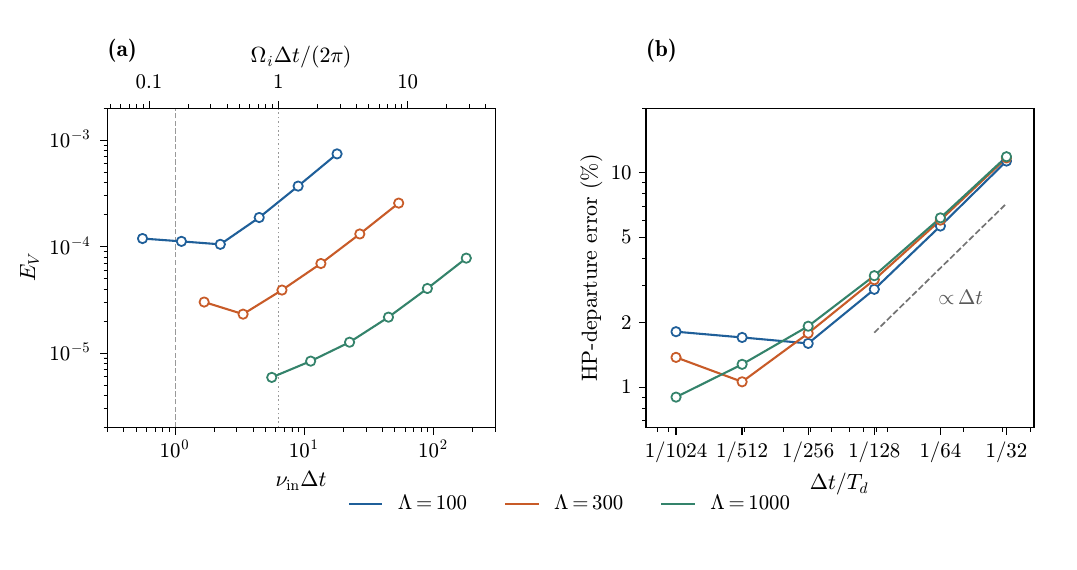}
  \caption{Large-step accuracy of the driven response at
  $\nuin=\Om=\Lambda$. (a)~Relative transfer difference $E_V$, with the upper axis giving
  gyroperiods per step. The vertical dashed and dotted
  lines mark $\nuin\dt=1$ and $\Om\dt=2\pi$, respectively.
  (b)~Relative error in the finite-frequency departure from the
  Hall--Pedersen values,
  $\|\Lambda(\bm H_{\PH}-\bm H_V)\|_2/
  \|\Lambda\bm H_V-\bm H_{HP}\|_2$.
  The dashed segment is proportional to $\dt$.}
  \label{fig:pathb-steps}
\end{figure}

In the Hall--Pedersen scaling the nonequilibrium stages are slaved to the
residual through $L_h^{-1}Q_h$, whose norm scales with $\nuin^{-1}$, and
the density update tends to the compact HP--Heun scheme
(\ref{app:proof-hp-theorem}). Motivated by this limiting behavior, the
driven response is assessed over time steps that are no longer small
compared with the collision time or the gyroperiod: each case
$\Lambda\in\{100,300,1000\}$ is repeated with $P\in\{512,256,128,64,32\}$ steps
per drive period (\cref{fig:pathb-steps}).
The density and nonequilibrium norm remain bounded throughout the $39$
drive periods, and $E_V<8\times10^{-4}$ for all fifteen cases.
At $\Lambda=1000$ and $P=32$, each step spans $178.5$ collision times and
$28.4$ gyroperiods, with $\vperp^{\max}\dt/\Delta x=7.7$, while
$E_V=7.8\times10^{-5}$.

The same coefficient difference is also measured relative to the
finite-frequency departure of the Volterra response from the
Hall--Pedersen values (\cref{fig:pathb-steps}(b)). This relative error is
below $2\%$ for $P=1024$, $512$ and $256$ and below $3.4\%$, $6.2\%$ and
$12\%$ for $P=128$, $64$ and $32$, respectively, and grows approximately
linearly with $\dt$ over the three coarsest step sizes.

\subsection{Hall--Pedersen density limit}
\label{sec:res-ap-packet}

The preceding tests concern the drive-frequency response. Here the
prescribed-field case of \Cref{prop:HP} is examined at fixed spatial and
velocity resolution and fixed slow-time step.
With the uniform acceleration $\bm a_0$ from \cref{sec:res-accuracy},
the $\mathcal K_6$ packet defined in \cref{eq:ap-packet} evolves under the scaling
\begin{equation}
\nuin=\varepsilon^{-1},\qquad
\Om=5\varepsilon^{-1},\qquad
\tau=\varepsilon t,\qquad
\dt=\frac{\Delta\tau}{\varepsilon}.
\label{eq:ap-scaling}
\end{equation}
The initial state is a Maxwellian plus a pure dipole,
\begin{equation}
\begin{gathered}
f_{0,h}^{\varepsilon}=n_{0,h}M_h+r_{h,\varepsilon}^{\mathrm{HP}},\\
(r_{h,\varepsilon}^{\mathrm{HP}})_{+1}
=\frac{(J_{x,h}-\mathrm iJ_{y,h})\vperp M_h}{2\theta_h},\\
(r_{h,\varepsilon}^{\mathrm{HP}})_{-1}
=\overline{(r_{h,\varepsilon}^{\mathrm{HP}})_{+1}},\qquad
(r_{h,\varepsilon}^{\mathrm{HP}})_m=0\quad(|m|\ne1).
\end{gathered}
\label{eq:ap-initial}
\end{equation}
Here $\bm J_h$ is the cell-centered Hall--Pedersen flux in
\cref{eq:cell-hp-flux}, evaluated at $n_{0,h}$ and $\bm a_0$, with the
discrete Maxwellian second moment $\theta_h$.
Since $\bm J_h=\Ord(\varepsilon)$, this initialization satisfies the
well-preparedness condition of \Cref{prop:HP}.

\FloatBarrier
The spatial--velocity grid is
$(N_x,N_y,N_{\vperp},M,N_\gyro)=(48,48,64,16,64)$ with
$\vperp^{\max}=8$.  We use $\Delta\tau=0.16$, $\tau_f=8$, and
\begin{equation*}
\varepsilon^{-1}\in\{160,240,320,480,640,800,960,1280\}.
\end{equation*}

\begin{figure}[!htbp]
  \centering
  \includegraphics[width=0.88200\textwidth]{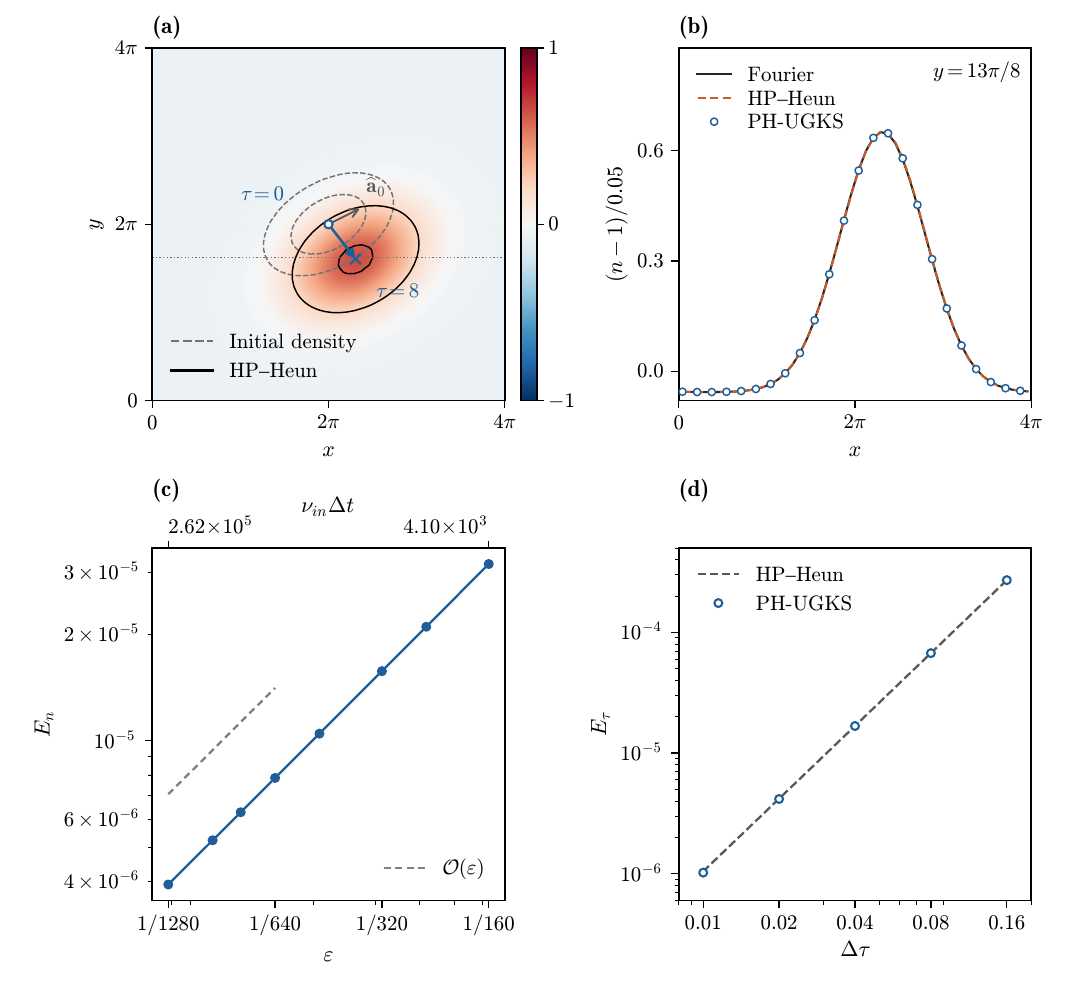}
  \caption{Hall--Pedersen drift, diffusion, and density limit.
  Panels (a,b) use $\varepsilon=1/160$ at $\tau=8$.
  (a)~$(n_h^\varepsilon-1)/0.05$, with initial-density contours
  (gray dashed) and HP--Heun contours (black solid) at the common levels
  $(n-1)/0.05=0.2,0.6$.
  The blue arrow shows the predicted displacement $8\widetilde{\bm u}$, and
  the gray arrow gives the direction of $\bm a_0$.
  (b)~Profiles along the horizontal dotted line $y=13\pi/8$: PH-UGKS,
  the same-grid HP--Heun solution, and the continuous Fourier reference.
  (c)~Density difference $E_n$ from HP--Heun for eight values of $\varepsilon$ at fixed
  spatial--velocity grid and $\Delta\tau=0.16$. The dashed segment is
  proportional to $\varepsilon$.
  (d)~Slow-time refinement at $\varepsilon=1/81920$. Both errors are
  measured against the exact-time solution of the compact Hall--Pedersen
  spatial discretization; the dashed curve is the HP--Heun time error.}
  \label{fig:ap-packet}
\end{figure}

The compact HP--Heun solution $N_h$ of \cref{sec:hp-property} uses the same
initial density, spatial grid, discrete transport coefficients, and
slow-time step.
The chosen slow-time step satisfies the stability bound obtained by
Fourier analysis of the HP--Heun scheme on this grid.
The normalized difference between the PH-UGKS and HP--Heun densities is
\begin{equation}
E_n(\varepsilon)=
\frac{\|n_h^\varepsilon(\tau_f)-N_h(\tau_f)\|_{2,h}}
{\|N_h(\tau_f)-1\|_{2,h}},
\label{eq:ap-error}
\end{equation}
where the density norm is the uniform-grid spatial RMS norm.

\begin{samepage}
For the prescribed uniform acceleration, the continuum drift velocity in
slow time is
\begin{equation*}
\widetilde{\bm u}
=\frac{\bm a_0-5\bb\times\bm a_0}{26}
=\frac{(7,-9)}{26\sqrt5}.
\end{equation*}
At $\tau=8$, the predicted displacement is $(0.963,-1.238)$.

\end{samepage}

\begin{samepage}
Hall drift deflects the packet from the acceleration direction, while Pedersen
diffusion, with $\widetilde D_{P,h}=\theta_h/26$, broadens the packet.
The continuous Fourier reference is the exact solution of the
constant-coefficient limit equation \eqref{eq:hp-density-limit} with
$\theta_n$ replaced by $\theta_h$, sampled by exact spatial cell averaging.

\end{samepage}

After 50 macroscopic steps, $E_n$ decreases from $3.16\times10^{-5}$ at
$\varepsilon=1/160$ to $3.92\times10^{-6}$ at $\varepsilon=1/1280$,
with a fitted slope of $1.00$ over all eight values
(\cref{fig:ap-packet}(c)), consistent with the finite-step density
estimate in \cref{sec:hp-property}. The slope persists at
$\Delta\tau=0.125$ and on the finer velocity grid
$(N_{\vperp},M,N_\gyro)=(96,20,80)$. At $\varepsilon=1/160$, each physical time step
spans 4096 collision times and approximately $3.26\times10^3$ gyroperiods.
The HP--Heun solution differs from the continuous Fourier reference by
$0.583\%$, normalized by the norm of the reference density perturbation,
and this difference decreases at second order under spatial refinement
with $\Delta\tau\propto\Delta x^2$ (observed orders $1.93$--$1.98$
between $64^2$ and $192^2$). This discretization error of the limiting scheme is distinct from $E_n$,
which compares the PH-UGKS and HP--Heun densities on the same grid.

Temporal refinement at $\varepsilon=1/81920$ examines the density evolution
on the slow time scale. This test uses
$(N_x,N_y,N_{\vperp},M,N_\gyro)=(48,48,80,16,64)$,
$\vperp^{\max}=10$, and the initial state in \cref{eq:ap-initial}, with
$\Delta\tau\in\{0.16,0.08,0.04,0.02,0.01\}$ and $\tau_f=8$.
The reference $N_h^{\mathrm{ex}}$ is the exact-time solution of the compact
Hall--Pedersen spatial discretization on the same grid, obtained by
exponentiating its Fourier symbol. The normalized density error $E_\tau$
is defined by \eqref{eq:ap-error}, with $N_h$ replaced by
$N_h^{\mathrm{ex}}$. It follows the second-order time error of HP--Heun,
with a fitted rate of $2.01$ (\cref{fig:ap-packet}(d)).

\subsection{Finite-amplitude self-consistent dynamics}
\label{sec:res-nonlinear}

The final test examines how ion--neutral collisions modify spatial-mode
generation and the mean-flux response in a finite-amplitude,
self-consistent electrostatic problem. We solve
\begin{equation}
\begin{gathered}
\partial_t f+\bv_\perp\cdot\grad_\perp f
+\left[\bm a_{\mathrm{ext}}(t)-\grad_\perp\phi
+\Om\bv_\perp\times\bb\right]\cdot\grad_{\bv}f
=\nuin(nM_n-f),\\
-\grad_\perp^2\phi=n-\langle n\rangle,
\qquad\langle\phi\rangle=0,
\end{gathered}
\label{eq:pulse-model}
\end{equation}
on the periodic domain $[0,2\pi]^2$, with angle brackets denoting spatial
averages. The initial density contains two noncollinear modes,
\begin{equation}
n_0(x,y)=1+0.16\left[\cos(x+y)
+0.6\cos\left(x-y+\frac{\pi}{4}\right)\right],
\qquad f_0=n_0M_n.
\label{eq:pulse-initial}
\end{equation}
The initial state is represented by exact spatial cell averages.

\FloatBarrier
A spatially uniform pulse drives the mean ion motion:
\begin{equation}
\begin{gathered}
\bm a_{\mathrm{ext}}(t)=0.15g(t)\widehat{\bm e}_a,
\qquad
\widehat{\bm e}_a=\frac{2\bm e_x+\bm e_y}{\sqrt5},\\
g(t)=
\begin{cases}
\sin^4(\pi t/T_p),&0\leq t\leq T_p,\\
0,&t>T_p,
\end{cases}
\qquad T_p=2\pi.
\end{gathered}
\label{eq:pulse-drive}
\end{equation}
We compute the response up to $t=4\pi$ for $\nuin=0,0.3,1,5$ at
$\Om=1$, with $k_\perp\rho_i=\sqrt2$ for both initial modes.
The discretization uses
$(N_x,N_y,N_{\vperp},M,N_\gyro)=(96,96,48,16,64)$,
$\vperp^{\max}=8$, and a spatial Courant number
$\vperp^{\max}\dt/\Delta x=0.375$.

\begin{figure}[!htbp]
  \centering
  \includegraphics[width=0.88200\textwidth]{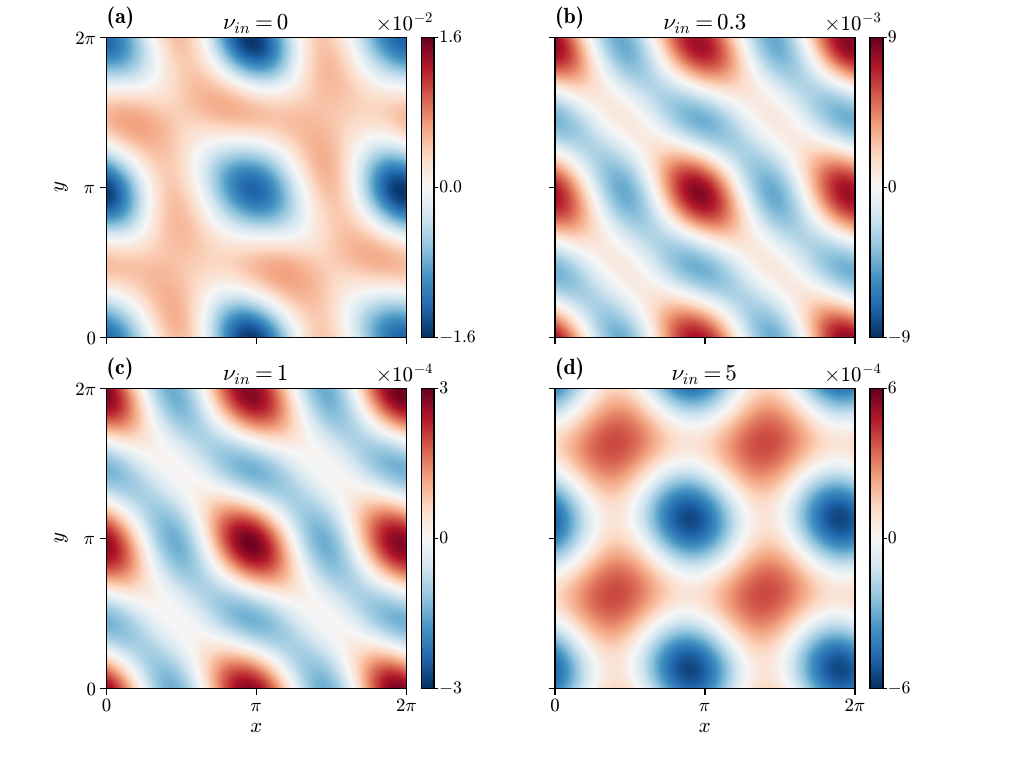}
  \caption{Density perturbation outside the initial spatial Fourier support,
  $\delta n_{\mathrm{out}}/0.16$, at pulse termination, $t=2\pi$.
  Panels (a)--(d) correspond to $\nuin=0,0.3,1,5$, respectively.}
  \label{fig:pulse-generation}
\end{figure}

\begin{figure}[!htbp]
  \centering
  \includegraphics[width=0.88200\textwidth]{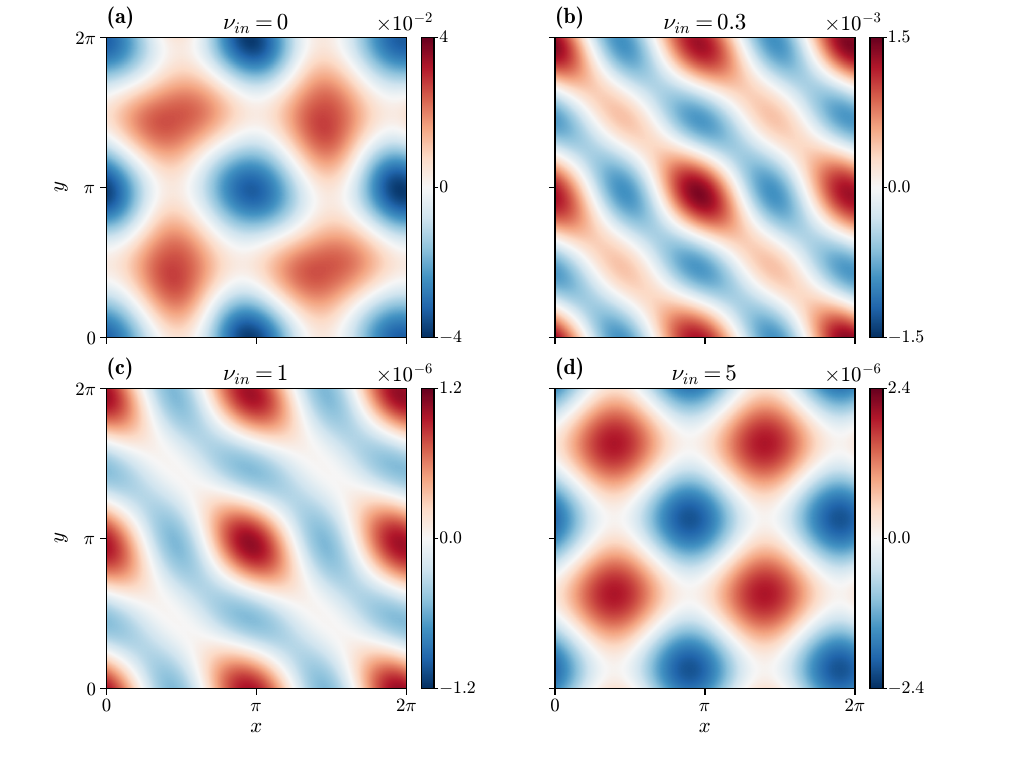}
  \caption{Density perturbation outside the initial spatial Fourier support,
  $\delta n_{\mathrm{out}}/0.16$, at $t=4\pi$, one gyroperiod after
  pulse termination. Panels (a)--(d) correspond to $\nuin=0,0.3,1,5$,
  respectively.}
  \label{fig:pulse-generation-final}
\end{figure}

With a prescribed uniform acceleration, the discrete update does not
couple distinct spatial Fourier modes. The self-consistent force instead
couples them through the product $-\grad_\perp\phi\cdot\grad_{\bv}f$.
The nonuniform initial state is not a self-consistent equilibrium.
In the continuous collisionless problem, the pulse can be removed in a
frame following the mean ion motion, so it changes the density Fourier
phases but not their amplitude histories.
To isolate the resulting spatial structure, we define
\begin{equation}
\mathcal S_0=\{\pm(1,1),\pm(1,-1)\},
\qquad
\delta n_{\mathrm{out}}=(I-P_{\mathcal S_0})(n-\langle n\rangle),
\label{eq:pulse-support}
\end{equation}
where $P_{\mathcal S_0}$ is the Fourier projection onto the initial
support. \Cref{fig:pulse-generation,fig:pulse-generation-final} show
$\delta n_{\mathrm{out}}/0.16$ at pulse termination and one gyroperiod
later. Collisions change both the amplitude and the spatial composition
of the generated perturbation.

At $t=4\pi$ (\cref{fig:pulse-generation-final}), the $\pm(2,2)$ pair, the
second spatial harmonic of the initially stronger mode, accounts for
$74.1\%$ of the discrete spatial $L^2$ norm squared of
$\delta n_{\mathrm{out}}$ for $\nuin=0.3$, producing the diagonal pattern.
For $\nuin=1$, this pair contributes $51.6\%$, while the two axial pairs
$\pm(2,0)$ and $\pm(0,2)$ together contribute $42.8\%$, giving a mixed
pattern. The axial pairs account for $96.3\%$ and $99.9\%$ at
$\nuin=0$ and $5$, respectively.

\begin{figure}[!htbp]
  \centering
  \includegraphics[width=0.88200\textwidth]{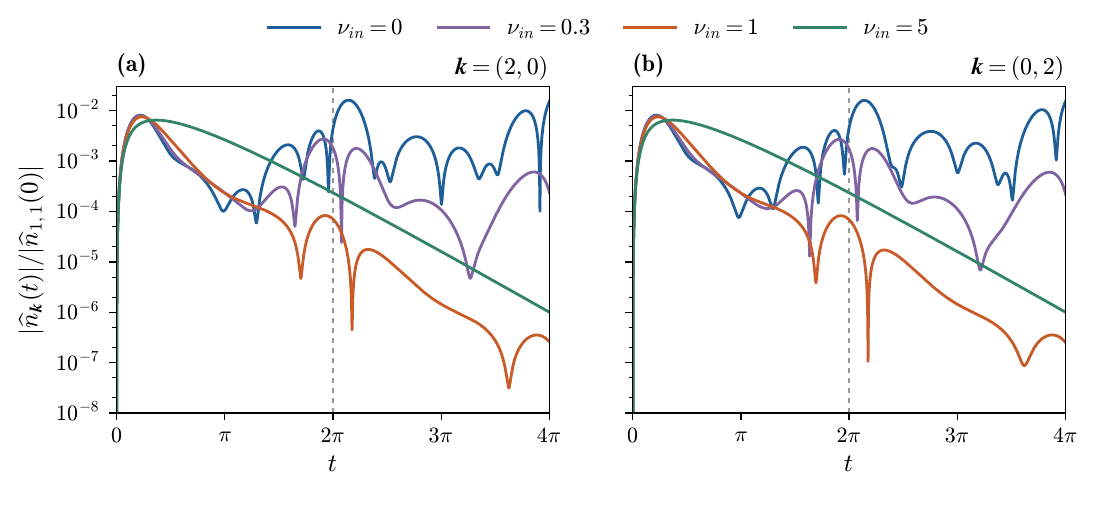}
  \caption{Evolution of the generated density Fourier amplitudes,
  computed by PH-UGKS and normalized by $|\widehat n_{1,1}(0)|$, for
  $\nuin=0,0.3,1,5$. Panels (a) and (b) show $\bm k=(2,0)$ and $(0,2)$,
  respectively. The vertical dashed lines mark pulse termination at
  $t=2\pi$.}
  \label{fig:pulse-modes}
\end{figure}

\Cref{fig:pulse-modes} shows the evolution of the $(2,0)$ and $(0,2)$
density modes, whose wavevectors are the sum and difference of the two
initial wavevectors. Their amplitudes are normalized by
$|\widehat n_{1,1}(0)|$. All four cases reach comparable peak amplitudes
during the pulse. After pulse termination, the collisionless response
remains oscillatory. The oscillations are weaker for $\nuin=0.3$, while
the amplitudes decay smoothly for $\nuin=5$. At $t=4\pi$, the smallest
normalized amplitudes, $(2.5\text{--}2.6)\times10^{-7}$, occur for
$\nuin=1$, compared with approximately $1.0\times10^{-6}$ for $\nuin=5$.

\begin{figure}[!htb]
  \centering
  \includegraphics[width=0.88200\textwidth]{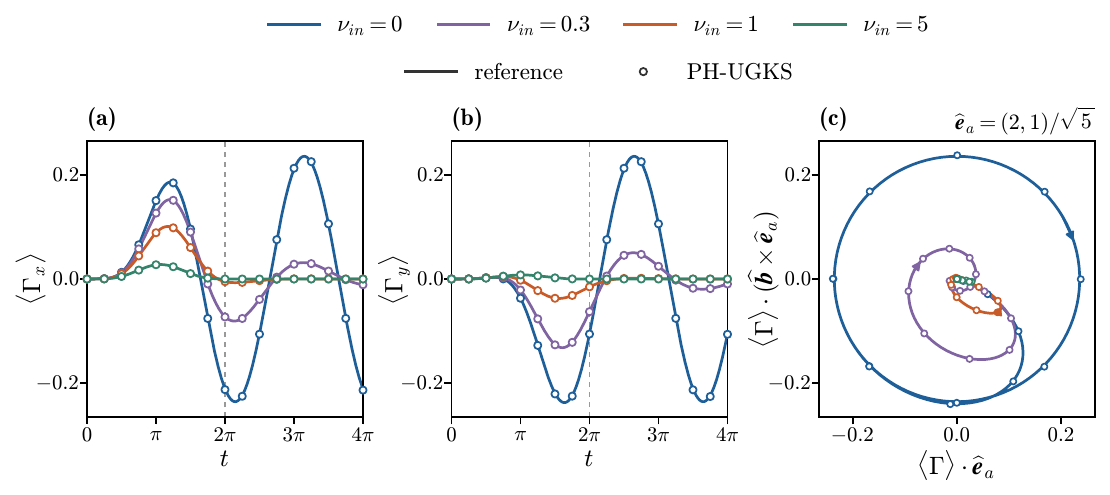}
  \caption{Spatially averaged ion-flux response for $\nuin=0,0.3,1,5$.
  Solid lines show the solution of \cref{eq:pulse-moment}, and symbols
  show PH-UGKS. Panels (a,b) show the Cartesian
  components; panel (c) shows the trajectory in the drive-aligned and
  transverse directions, with $\widehat{\bm e}_a=(2,1)/\sqrt5$.
  Arrows in (c) indicate increasing time. Vertical dashed lines in (a,b) mark
  pulse termination.}
  \label{fig:pulse-mean-flux}
\end{figure}

For the continuous periodic Poisson problem, the mean internal force
density vanishes, $\langle n(-\grad_\perp\phi)\rangle=\bm0$, providing
an independent reference for the spatially averaged ion flux.
With $\langle n\rangle=1$, the first-moment equation gives
\begin{equation}
\frac{\mathrm d\langle\bm\Gamma_\perp\rangle}{\mathrm dt}
+\nuin\langle\bm\Gamma_\perp\rangle
-\Om\langle\bm\Gamma_\perp\rangle\times\bb
=\bm a_{\mathrm{ext}}(t),
\qquad \langle\bm\Gamma_\perp\rangle(0)=\bm0.
\label{eq:pulse-moment}
\end{equation}
Once the pulse ends, the reference flux rotates at $\Om$, with magnitude
proportional to $\exp[-\nuin(t-T_p)]$. Thus the collisionless trajectory
completes a circle during $T_p\leq t\leq4\pi$, whereas collisional
relaxation contracts the trajectory towards the origin.
\Cref{fig:pulse-mean-flux} compares the numerical and reference fluxes
in Cartesian components and in drive-aligned coordinates.
The mean ion flux agrees with the collision--rotation reference, while
collisions alter both the amplitudes and spatial composition of the
generated density modes.

\section{Conclusions}\label{sec:conclusion}

PH-UGKS advances the ion distribution through a conservative density
update and an exponential update of its nonequilibrium component.
The method combines exact collision--rotation integration, a kinetic flux
reconstructed from the Duhamel time average, and a compact Hall--Pedersen
correction to the density flux.
It conserves discrete ion number on periodic domains and is second-order
consistent in time for fixed collision and gyrofrequencies.
For well-prepared perpendicular transport under the stated assumptions,
the analysis establishes the Hall--Pedersen density limit as collisionality
increases at fixed magnetization, spatial--velocity grids, and slow-time step.

\FloatBarrier

Collisionless benchmarks reproduce ion Bernstein dispersion and
Dory--Guest--Harris growth rates.
In the driven tests at fixed $k_\perp\rho_i$, increasing the
collision-to-gyrofrequency ratio concentrates the nongyrotropic response
in the dipolar harmonics.
In the finite-amplitude Poisson test, ion--neutral collisions modify the
amplitudes and spatial composition of density modes generated outside
the initial Fourier support. The spatially averaged ion flux agrees
with the corresponding collision--rotation moment equation.

In strongly collisional driven tests, the ion-flux response agrees with
the characteristic--Volterra reference at time steps far larger than both
the collision time and the gyroperiod, without subcycling.
The finite-frequency departure from the Hall--Pedersen response is also
resolved. Its leading term reflects ion inertia and pressure feedback
through the density response.
Over the coarsest tested time steps, the error in this departure decreases
approximately linearly with the time step.
Fixed-resolution density tests reproduce Hall drift and Pedersen diffusion
and confirm convergence to the same-grid compact Hall--Pedersen--Heun
solution. These results show that the same kinetic discretization retains
finite-frequency response and recovers macroscopic transport without
switching to a fluid solver.

The present model assumes a uniform magnetic field and constant-frequency
BGK relaxation toward a stationary neutral Maxwellian.
Future work will consider nonuniform magnetic fields and more general
ion--neutral collision operators.

\section*{Declaration of competing interest}

The authors declare that they have no known competing financial interests or
personal relationships that could have appeared to influence the work reported
in this paper.

\section*{Acknowledgements}

The current research is supported by the National Key R\&D Program of China
(Grant No. 2022YFA1004500), the National Natural Science Foundation of China
(92371107), and the Hong Kong Research Grants Council (16208324).

\clearpage
\appendix
\Needspace{10\baselineskip}
\section{Hall--Pedersen limit of the density update}
\label{app:hp-proof}

\subsection{Hall--Pedersen moment relation}
\label{app:discrete-hp-moment}

The complete velocity operator is assumed to conserve ion number in
each spatial cell and to satisfy
\[
\left\langle
\boldsymbol v_\perp D_{v,h}(nM_{n,h},\boldsymbol a)
\right\rangle_h
=-n\boldsymbol a_\perp.
\]

Let $L_h^{-1}$ denote the inverse of the restriction of $L_h$ to
$\operatorname{Ran}Q_h$. Define the discrete Chapman--Enskog state
\begin{equation}
\label{eq:hp-reference}
r_h(n,\boldsymbol a)
=
L_h^{-1}Q_h
\left.\mathcal R_h(nM_{n,h},\boldsymbol a)\right|_{T_3=0}.
\end{equation}
For perpendicular dynamics at $f=nM_{n,h}$, the interior force in
\crefrange{eq:polar-force}{eq:velocity-force-flux} and the spatial flux
at $T_3=0$ contain only the harmonics $m=\pm1$. The analysis assumes
that the complete residual retains this harmonic support:
\[
\left[
 Q_h\left.\mathcal R_h(nM_{n,h},\boldsymbol a_\perp)\right|_{T_3=0}
\right]_m=0,
\qquad m\notin\{-1,1\}.
\]
The force moment above and the spatial flux at $T_3=0$ give
\[
\left\langle
\boldsymbol v_\perp Q_h
\left.\mathcal R_h(nM_{n,h},\boldsymbol a)\right|_{T_3=0}
\right\rangle_h
=
n\boldsymbol a_\perp-\theta_h\nabla_{\perp,h}n.
\]

With $\boldsymbol J_h=\langle\boldsymbol v_\perp r_h\rangle_h$ and
$\boldsymbol d_h=n\boldsymbol a_\perp-\theta_h\nabla_{\perp,h}n$,
the identity
$\langle\boldsymbol v_\perp D_{\gyro,h}r_h\rangle_h
=\boldsymbol J_h\times\bb$ gives
\begin{equation}
\label{eq:discrete-hp-balance}
\nuin\boldsymbol J_h-\Om\boldsymbol J_h\times\bb
=\boldsymbol d_h.
\end{equation}
Solving \eqref{eq:discrete-hp-balance} recovers the cell-centered
Hall--Pedersen flux in \cref{eq:cell-hp-flux}.

Since $L_h^{-1}$ preserves each gyroharmonic, $r_h$ has support only in
$m=\pm1$. Under $\gyro\mapsto\gyro+\pi$, $r_h$ changes
sign and $|v_n|$ is unchanged. The equally spaced gyroangle grid with an
even number of nodes therefore gives
\[
\left\langle
|v_n|(r_{h,F,L}-r_{h,F,R})
\right\rangle_h=0.
\]
Using \cref{eq:angle-operators}, we obtain
\begin{equation}
\label{eq:cell-face-hp-moment}
\left\langle
\mathcal V_{F,h}(r_{h,F,L},r_{h,F,R})
\right\rangle_h
=
\left\{\boldsymbol J_h\cdot\bn_F\right\}_F
=
\Gamma_{F,h}^{\mathrm{cell}}.
\end{equation}

\subsection{Proof of Theorem~\ref{prop:HP}}
\label{app:proof-hp-theorem}

The symbol $\|\cdot\|_h$ also denotes the induced operator norm. All
estimates below are uniform for $0<\varepsilon\le\varepsilon_0$. On
$\operatorname{Ran}Q_h$,
\[
L_h
=\varepsilon^{-1}(I-\beta D_{\gyro,h}),
\qquad
\dt L_h
=\frac{\Delta\tau}{\varepsilon^2}(I-\beta D_{\gyro,h}).
\]
The eigenvalues of $I-\beta D_{\gyro,h}$ are $1$ on the non-Maxwellian part
of the zeroth gyroharmonic and $1-\mathrm i m\beta$ on harmonic $m\ne0$.
Consequently,
\[
\|L_h^{-1}Q_h\|_h\le C\varepsilon,
\qquad
\|S_{\dt}Q_h\|_h
\le C\exp\!\left(-\frac{\Delta\tau}{\varepsilon^2}\right).
\]
The definitions in \cref{eq:phis} give
\begin{align*}
\dt\Phi_0(\dt L_h)Q_h
&=L_h^{-1}Q_h
 +\mathcal O\!\left(\varepsilon
 \mathrm e^{-\Delta\tau/\varepsilon^2}\right),
&
\Phi_0(\dt L_h)Q_h
&=\mathcal O(\varepsilon^2),
\\
\dt(\Phi_1-\Phi_2)(\dt L_h)Q_h
&=\frac12L_h^{-1}Q_h+\mathcal O(\varepsilon^3),
&
\dt\Phi_2(\dt L_h)Q_h
&=\frac12L_h^{-1}Q_h+\mathcal O(\varepsilon^3).
\end{align*}
Moreover,
$T_3(\Delta\tau/\varepsilon^2)=o(\varepsilon^r)$ for every $r>0$.

By \cref{eq:ph-flux}, setting $T_3=0$ changes $\mathcal R_h$ by
$\mathcal O(T_3)$ on the fixed grid and the prescribed bounded sets.
The Lipschitz bound and
$f^n-n^nM_{n,h}=\mathcal O(\varepsilon)$ give
\[
Q_h R^0
=
Q_h\left.\mathcal R_h(n^n M_{n,h},\boldsymbol a^0)\right|_{T_3=0}
+\mathcal O(\varepsilon).
\]
Substitution into \cref{eq:phugks-update-b} yields
\[
h^{(1)}
=
r_h(n^n,\boldsymbol a^0)+\mathcal O(\varepsilon^2).
\]
Equations~\eqref{eq:ph-flux} and \eqref{eq:cell-face-hp-moment} then give
\[
\Gamma_{F,h}^{\PH}\!\left(n^n M_{n,h}+h^{(1)}\right)
=
\Gamma_{F,h}^{\mathrm{cell}}(n^n,\boldsymbol a^0)
-T_3(\nuin\dt)d_{n,h}[n^n]_F
+\mathcal O(\varepsilon^2).
\]
Adding \cref{eq:hp-correction} and using the decay of $T_3$ gives
\[
\begin{aligned}
\Gamma_{F,h}^{\PH}\!\left(n^n M_{n,h}+h^{(1)}\right)
+\Gamma_{F,h}^{\mathrm{corr},0}
&=
\Gamma_{F,h}^{\mathrm{cmp}}(n^n,\boldsymbol a^0)
+\mathcal O(\varepsilon^2)+\mathcal O(T_3)
\\
&=
\Gamma_{F,h}^{\mathrm{cmp}}(n^n,\boldsymbol a^0)
+\mathcal O(\varepsilon^2).
\end{aligned}
\]
Since
$\Gamma_{F,h}^{\mathrm{cmp}}
=\varepsilon\widetilde\Gamma_{F,h}^{\mathrm{cmp}}$
and $\dt=\Delta\tau/\varepsilon$, substitution into
\cref{eq:phugks-update-c} proves \cref{eq:HP-predictor}.

The predictor estimate gives $h^{(1)}=\mathcal O(\varepsilon)$ and
$n^{(1)}=\mathcal O(1)$. The same argument at the second stage gives
\[
Q_h R^1
=
Q_h\left.\mathcal R_h(n^{(1)} M_{n,h},\boldsymbol a^1)\right|_{T_3=0}
+\mathcal O(\varepsilon).
\]
Applying $Q_h$ to \cref{eq:time-averaged-distribution} yields
\[
Q_h\bar f
=
\frac12\left[
r_h(n^n,\boldsymbol a^0)
+r_h(n^{(1)},\boldsymbol a^1)
\right]
+\mathcal O(\varepsilon^2).
\]
Since $h^n,h^{(1)}=\mathcal O(\varepsilon)$ and $T_3=o(\varepsilon)$,
the number-flux formula following \cref{eq:ph-flux} and ion-number
conservation of $D_{v,h}$ give
$\langle R^0\rangle_h,\langle R^1\rangle_h=\mathcal O(\varepsilon)$.
Taking the density moment of \cref{eq:time-averaged-distribution} yields
\[
\bar n:=\langle\bar f\rangle_h
=n^n+\dt\left(\frac13\langle R^0\rangle_h
              +\frac16\langle R^1\rangle_h\right)=\mathcal O(1).
\]
The Maxwellian contribution to its spatial number flux is
$-T_3(\chi)d_{n,h}[\bar n]_F=\mathcal O(T_3)$. With the stage corrections,
the numerical number flux therefore satisfies
\[
\Gamma_{F,h}^{\PH}(\bar f)
+\frac12\left(
\Gamma_{F,h}^{\mathrm{corr},0}
+\Gamma_{F,h}^{\mathrm{corr},1}
\right)
=
\frac12\left[
\Gamma_{F,h}^{\mathrm{cmp}}(n^n,\boldsymbol a^0)
+\Gamma_{F,h}^{\mathrm{cmp}}(n^{(1)},\boldsymbol a^1)
\right]
+\mathcal O(\varepsilon^2).
\]
Substitution into \cref{eq:density-update} proves
\cref{eq:HP-estimate}.

It remains to estimate the nonequilibrium component at the new time level.
The definitions in \cref{eq:phis} give
\[
\dt(\Phi_0-\Phi_1)(\dt L_h)Q_h
=\mathcal O(\varepsilon^3),
\qquad
\dt\Phi_1(\dt L_h)Q_h
=L_h^{-1}Q_h+\mathcal O(\varepsilon^3).
\]
Equation~\eqref{eq:intermediate-distribution} gives
$Q_h\widetilde f=\mathcal O(\varepsilon)$. The uniform Lipschitz assumption
in \Cref{prop:HP} gives $R^\star-R^1=\mathcal O(1)$. Applying $Q_h$ to
\cref{eq:final-update} yields
\[
\|Q_h f^{n+1}\|_h=\mathcal O(\varepsilon),
\]
which completes the one-step estimate.

For the finite-step conclusion, let $\mathcal H_j$ be the compact
Hall--Pedersen--Heun map at slow time $\tau^j$. The predictor formula
\eqref{eq:HP-predictor} differs by $\mathcal O(\varepsilon)$ from its
limiting Euler stage. The rescaled compact flux is locally Lipschitz in the
density and field values on bounded sets. Equation~\eqref{eq:HP-estimate}
therefore gives
\[
n_{\varepsilon,h}^{j+1}
=\mathcal H_j(n_{\varepsilon,h}^j)+\mathcal O(\varepsilon),
\qquad
N_h^{j+1}=\mathcal H_j(N_h^j).
\]
For $e_j=\|n_{\varepsilon,h}^j-N_h^j\|_\infty$, the uniform Lipschitz
bound gives $e_{j+1}\le L_{\mathcal H}e_j+C\varepsilon$.
Since $N$ is fixed independently of $\varepsilon$, iteration yields
$\max_{0\le j\le N}e_j\le C_N(e_0+\varepsilon)$.
The nonequilibrium bound follows by successive application of the
one-step estimate.

\FloatBarrier
\setcounter{table}{0}
\setcounter{figure}{0}
\makeatletter
\renewcommand{\theHtable}{app.\thesection.\arabic{table}}
\renewcommand{\theHfigure}{app.\thesection.\arabic{figure}}
\makeatother
\section{Additional collisionless tests}
\label[appendix]{app:collisionless-validation}

Complete potential histories and selected branch phases for the model in
\cref{eq:nibw-model} are compared with those from an independently
implemented backward semi-Lagrangian solver using the rotating-frame
formulation of Schild et al.~\cite{Schild2025}.
The reference uses seven-point Lagrange interpolation and Strang splitting:
a full spatial-advection step lies between two velocity-advection half steps,
with the electric field updated after spatial advection.

For $k_\perp\rho_i=0.5$, $1$, and $2$, PH-UGKS uses the grid
$(N_x,N_{\vperp},M,N_\gyro)=(32,80,28,64)$ with $\dt=0.01$ and $T=24$.
The rotating-grid reference uses
$N_x\times N_{v_x}\times N_{v_y}=48\times81\times81$ with the same time step
and final time. Both solvers start from the same continuous cosine perturbation.
The complex potential histories are normalized by
their respective initial Fourier coefficients. Across the three
wavenumbers, the relative $\ell^2$ differences from the rotating-grid
reference over $0\le t\le24$ are at most $2.49\%$.
Refining the PH-UGKS spatial grid to $N_x=64$, with $\dt=0.005$, reduces
the difference at $k_\perp\rho_i=2$ to $0.53\%$.

\begin{figure}[!ht]
  \centering
  \includegraphics[width=0.88200\textwidth]{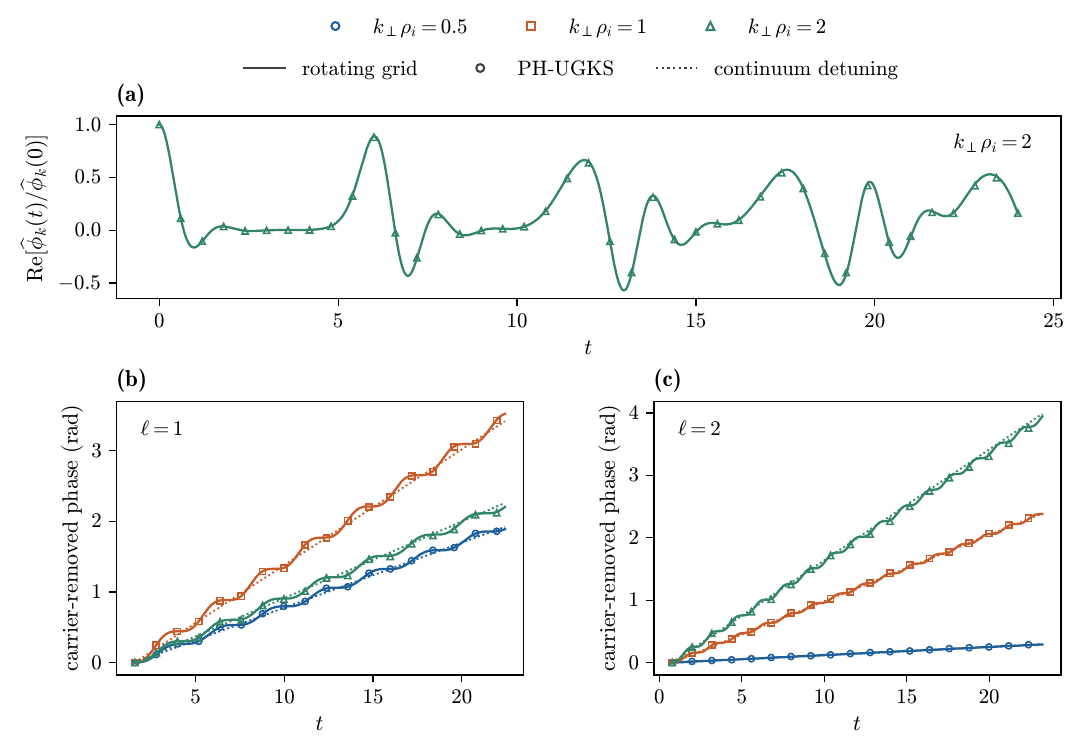}
  \caption{Comparison with the independently implemented
  rotating-grid solver. (a)~Real part of the complete spatial-mode potential history at
  $k_\perp\rho_i=2$.
  (b,c)~Carrier-removed phases of the extracted branches $\ell=1$ and
  $\ell=2$, respectively, at the three wavenumbers.
  Dotted lines have the slopes $\omega_{\ell,\mathrm{ref}}-\ell\Om$
  given by the continuum dispersion relation.}
  \label{fig:rotating-final}
\end{figure}

To resolve the phase evolution relative to cyclotron rotation, branches 1
and 2 are extracted from each complete history with the matrix-pencil
estimator of \cref{sec:res-kend}, demodulated by
$\mathrm e^{-\mathrm i\ell\Om t}$, and smoothed with a centered moving
average of width approximately $\pi/(\ell\Om)$ over complete averaging
windows only, with the unwrapped phase set to zero at the first retained
time. The resulting phases differ by at most $0.022$ rad
between the two solvers (\cref{fig:rotating-final}(b,c)).

\begin{samepage}
Separate spatial and velocity refinements of the rotating-grid reference
to $96\times81\times81$ and $48\times161\times161$, respectively, change
the complete normalized potential histories by less than $4\times10^{-6}$
in relative norm.

\end{samepage}

We next consider the Dory--Guest--Harris instability of a non-Maxwellian
ring equilibrium~\cite{Dory1965,Vogman2014}. The periodic
Vlasov--Poisson system is
\begin{equation}
\partial_t f+v_x\partial_x f+E_x\partial_{v_x}f-\Om\partial_\gyro f=0,
\qquad
-R^{-2}\partial_{xx}\phi=n-1,
\qquad E_x=-\partial_x\phi,
\label{eq:dgh-model}
\end{equation}
where $R^2=400$ and $\Om=1$.  The stationary equilibrium is
\begin{equation}
F_0(\vperp)=\frac{1}{\pi\alpha^2j!}
\left(\frac{\vperp^2}{\alpha^2}\right)^{j}
\mathrm e^{-\vperp^2/\alpha^2},
\qquad j=6,\qquad \alpha=\sqrt{2/j},
\label{eq:dgh-ring}
\end{equation}
whose maximum is at $\vperp=\sqrt2$.  Its unit-mass discrete cell averages
$F_{0,h}$ are perturbed by
\begin{equation}
f_h(x,\vperp,\gyro,0)=F_{0,h}(\vperp)
\left[1+10^{-4}\sin(4\gyro-kx)\right],
\qquad k=\widetilde k/\sqrt2.
\label{eq:dgh-seed}
\end{equation}
The radial force flux uses the fixed ring equilibrium $F_{0,h}$ as its
reference distribution (\cref{sec:method-interface}).

Each spatial domain contains one wavelength.  The stable case
$\widetilde k=0.5$ uses $(N_x,N_{\vperp})=(24,64)$, and the purely growing
case $\widetilde k=3.15$ uses $(48,128)$. Both end at $T=18$.
The oscillatory growing case $\widetilde k=4.65$ uses
$(N_x,N_{\vperp})=(128,256)$ and $T=25.5$.
All three cases use $(M,N_\gyro)=(12,52)$, $\vperp^{\max}=4$, and a spatial
Courant number $\vperp^{\max}\dt/\Delta x\simeq0.3$.

The reference roots are zeros of the electrostatic dispersion
function~\cite{Dory1965,Vogman2014}.  With the convention
$\mathrm e^{-\mathrm i\omega t}$, they give
\begin{equation}
\gamma=0.491198
\quad(\widetilde k=3.15),
\qquad
\omega=1.036093+0.289857\,\mathrm i
\quad(\widetilde k=4.65).
\label{eq:dgh-roots}
\end{equation}
With $U_E=(\Delta x/2)\sum_i E_i^2$, the numerical growth rates $\gamma_h$
are half the least-squares slope of $\ln U_E$, over $[7.5,15]$ for the
purely growing mode and through six maxima in $[6.9,25.5]$ for the
oscillatory mode, whose fitted spacing $\Delta t_{\mathrm{pk}}$ gives
$\omega_{r,h}=\pi/\Delta t_{\mathrm{pk}}$.

\begin{figure}[!htbp]
  \centering
  \includegraphics[width=0.88200\textwidth]{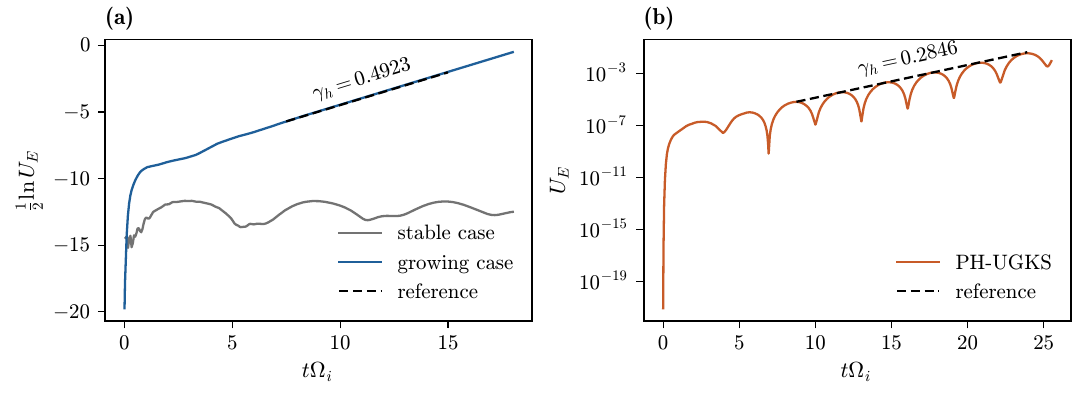}
  \caption{Dory--Guest--Harris benchmark. (a)~$\tfrac12\ln U_E$ for the
  stable ($\widetilde k=0.5$) and purely growing ($\widetilde k=3.15$) cases.
  (b)~Electric-field energy for the refined oscillatory growing case
  ($\widetilde k=4.65$). Dashed lines show reference growth envelopes with
  dispersion-relation growth rates, matched in amplitude at the first
  fitted time.}
  \label{fig:dgh-final}
\end{figure}

The stable case shows no sustained exponential growth. Relative to the
dispersion-relation roots, the growth-rate differences are $0.228\%$ for
the purely growing mode and $1.82\%$ for the oscillatory mode, for which
the real-frequency difference is $0.018\%$
(\cref{fig:dgh-final}).

\begin{figure}[!htbp]
  \centering
  \includegraphics[width=0.88200\textwidth]{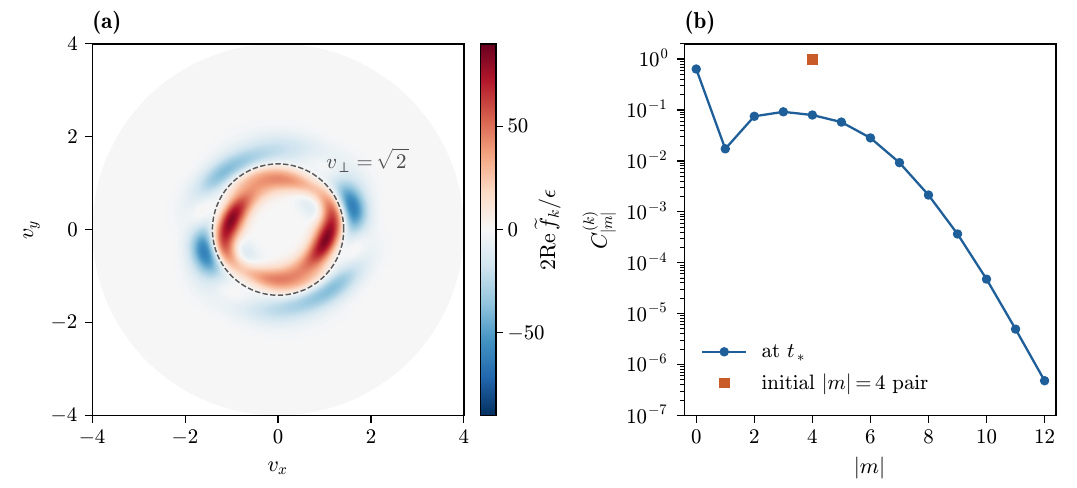}
  \caption{Velocity-space structure of the evolving Dory--Guest--Harris
  perturbation at $\widetilde k=4.65$ and $t_\ast=17.84$, from the
  $(N_x,N_{\vperp})=(64,128)$ solution.
  (a)~$2\operatorname{Re}\widetilde f_k/\epsilon$ for the fundamental
  spatial Fourier coefficient of $f_h-F_{0,h}$, with $\epsilon=10^{-4}$
  and phase chosen so that the density coefficient is real and positive.
  (b)~Paired gyroharmonic fractions $C^{(k)}_{|m|}$, normalized by the total
  content including $m=0$.}
  \label{fig:dgh-state-final}
\end{figure}

For the oscillatory case $\widetilde k=4.65$, the distribution is sampled
at the field-energy maximum $t_\ast=17.84$ on the grid
$(N_x,N_{\vperp})=(64,128)$, using the same Courant number and remaining
parameters. The frequency and growth-rate comparisons above use the finer
$(128,256)$ grid. Here $\widetilde f_k$ denotes the fundamental spatial
Fourier coefficient of $\delta f_h=f_h-F_{0,h}$, with its phase chosen so
that the density coefficient is real and positive, and the paired
gyroharmonic fractions are
$C^{(k)}_{|m|}=\mathcal E_{|m|}/\sum_{q=0}^{M}\mathcal E_q$,
with $\mathcal E_{|m|}$ evaluated on $\widetilde f_{k,m}$ using
\cref{eq:paired-content}.
The initial $|m|=4$ perturbation leaves the density uniform, so the
substantial $m=0$ component at $t_\ast$ (\cref{fig:dgh-state-final})
develops during the evolution and carries the density response, together
with several additional gyroharmonics. The perturbation remains
concentrated near the equilibrium ring, with rapidly decreasing
contributions at high $|m|$.

\FloatBarrier

\end{document}